\documentclass[a4paper,fleqn,usenatbib]{mnras}
\usepackage{graphicx}   
\usepackage{amsmath}    
\usepackage{amssymb}    
\usepackage{multicol}        
\usepackage{bm}     
\usepackage{pdflscape}  

\usepackage{multirow}
\usepackage{textcomp}
\usepackage{amssymb}

\usepackage[T1]{fontenc}
\usepackage{ae,aecompl}

\usepackage{hyperref}
\usepackage{caption}

\usepackage[utf8]{inputenc}

\newcommand{\cxo}{{\em Chandra}}

\newcommand{\xmm}{{\em XMM--Newton}}

\newcommand{\swift}{{\em Swift}}
\newcommand{\fermi}{{\em Fermi}}
\newcommand{\INT}{{\em INTEGRAL}}
\newcommand{\nus}{{\em NuSTAR}}

\def\nh {$N_{H}$}
\def\chisq {$\chi ^{2}$}
\def\rchisq {$\chi_{\nu} ^{2}$}
\def\lum {erg\,s$^{-1}$}
\def\flux {erg\,s$^{-1}$cm$^{-2}$}
\def\ss {s\,s$^{-1}$}
\def\cm2{cm$^{-2}$}
\def\arcsec{$^{\prime\prime}$}
\def\arcmin{$^{\prime}$}

\def\kt{$kT_{\rm BB}$}
\def\r{$R_{\rm BB}$}

\def\srclong{CXOU\,J164710.2--455216}
\def\src{CXOU\,J1647}

\graphicspath{{./plot/}}

\title[Multi-outburst activity of \src]{The multi-outburst activity of the magnetar in Westerlund I}

\author[A. Borghese et al.]{
A.~Borghese$^{1}$\thanks{E-mail: a.borghese@uva.nl}, N.~Rea$^{1,2,3}$, R.~Turolla$^{4,9}$, J.~A.~Pons$^{5}$, P.~Esposito$^{1,6}$, F.~Coti~Zelati$^{2,3}$, 
\newauthor 
V.~Savchenko$^{7}$, E.~Bozzo$^{7}$, R.~Perna$^{8}$, S.~Zane$^{9}$, S.~Mereghetti$^{6}$, S.~Campana$^{10}$, 
\newauthor 
R.~P.~Mignani$^{6,11}$, M.~Bachetti$^{12}$, G.~Rodr\'iguez$^{13}$, F.~Pintore$^{6}$, A.~Tiengo$^{6,14,15}$, D.~G\"{o}tz$^{16}$,
\newauthor
G.~L.~Israel$^{13}$, L.~Stella$^{13}$
\\
$^{1}$Anton Pannekoek Institute for Astronomy, University of Amsterdam, Postbus 94249, NL--1090 GE Amsterdam, The Netherlands\\
$^{2}$Institute of Space Sciences (ICE, CSIC), Campus UAB, Carrer de Can Magrans s/n, 08193 Barcelona, Spain\\
$^{3}$Institut d’Estudis Espacials de Catalunya (IEEC), 08034 Barcelona, Spain\\
$^{4}$Dipartimento di Fisica e Astronomia, Universit\`a di Padova, via F. Marzolo 8, I-35131 Padova, Italy\\
$^{5}$Departament de Fisica Aplicada, Universitat d’Alacant, Ap. Correus 99, E-03080 Alacant, Spain\\ 
$^{6}$INAF -- Istituto di Astrofisica Spaziale e Fisica Cosmica, via E. Bassini 15, I-20133 Milano, Italy\\
$^{7}$ISDC, Department of Astronomy, University of Geneva, Chemin d’Écogia 16, CH-1290 Versoix, Switzerland
$^{8}$Department of Physics and Astronomy, Stony Brook University, Stony Brook, NY 11794, USA\\
$^{9}$Mullard Space Science Laboratory, University College London, Holmbury St. Mary, Dorking, Surrey RH5 6NT, UK\\
$^{10}$INAF -- Osservatorio Astronomico di Brera, via Bianchi 46, I-23807 Merate (LC), Italy\\
$^{11}$Janusz Gil Institute of Astronomy, University of Zielona G\'ora, Lubuska 2, 65-265, Zielona G\'ora, Poland\\
$^{12}$INAF -- Osservatorio Astronomico di Cagliari, via della Scienza 5, I-09047 Selargius (CA), Italy\\
$^{13}$INAF -- Osservatorio Astronomico di Roma, via Frascati 33, I-00040 Monteporzio Catone (RM), Italy\\
$^{14}$Istituto Nazionale di Fisica Nucleare, Sezione di Pavia, via A. Bassi 6, I-27100 Pavia, Italy\\
$^{15}$Scuola Universitaria Superiore IUSS Pavia, piazza della Vittoria 15, I-27100 Pavia, Italy\\
$^{16}$AIM–CEA/DRF/Irfu/Service d’Astrophysique, Orme des Merisiers, F-91191 Gif-sur-Yvette, France\\
}

\date{Accepted XXX. Received YYY; in original form ZZZ}

\pubyear{2018}

\begin{document}
\label{firstpage}
\pagerange{\pageref{firstpage}--\pageref{lastpage}}
\maketitle

\begin{abstract}

After two major outbursts in 2006 and 2011, on 2017 May 16 the
magnetar \srclong, hosted within the massive star cluster
Westerlund I, emitted a short ($\sim$ 20~ms) burst, which marked
the onset of a new active phase. We started a  long-term
monitoring campaign with \swift\ (45 observations), \cxo\ (5
observations) and \nus\ (4 observations) from the activation until
2018 April. During the campaign, \swift\ BAT registered the
occurrence of multiple bursts, accompanied by two other
enhancements of the X-ray persistent flux. The long time span
covered by our observations allowed us to study the spectral as
well as the timing evolution of the source. After $\sim$ 11 months
since the 2017 May outburst onset, the observed flux was $\sim$ 15
times higher than its historical minimum level and a factor of
$\sim$ 3 higher than the level reached after the 2006 outburst.
This suggests that the crust has not fully relaxed to the
quiescent level, or that the source quiescent level has changed
following the multiple outburst activities in the
past 10 years or so. This is another case of multiple outbursts
from the same source on a yearly time scale, a somehow recently
discovered behaviour in magnetars.


\end{abstract}


\section{Introduction}


Since they were first discovered in 1979 \citep{1979Natur.282..587M},
Anomalous X-ray Pulsars (AXPs) and Soft Gamma-ray Repeaters (SGRs) have 
reached a total of 29 sources\footnote{See the online McGill Magnetar Catalog,
  http://www.physics.mcgill.ca/$\sim$pulsar/magnetar/main.html
  \citep{2014ApJS..212....6O}.}. Initially interpreted as two
different classes, it is now believed that there is no intrinsic
distinction, and they are cumulatively referred to as `magnetars', isolated neutron
stars (NSs) powered by ultra-strong magnetic fields \citep[see][for
  reviews]{2015RPPh...78k6901T, 2017ARA&A..55..261K,
  2018arXiv180305716E}. They display X-ray pulsations with periods
in the 0.3 -- 12~s interval\footnote{The source at the centre of the
  supernova remnant RCW\,103 is an exception, being the slowest
  magnetar ever detected with its 6.67-hr spin period
  \citep{2016ApJ...828L..13R}.} and relatively large spin-down rates
($\dot{P} \sim$ 10$^{-15}$ -- 10$^{-11}$~\ss). Assuming that they are
slowed down by magneto-rotational losses, the surface dipolar magnetic
field strength, as inferred from the timing properties, is as high as
$\sim$ 10$^{14}$ -- 10$^{15}$~G, with the exception of a handful of
objects that show a magnetic field in the range of those of the
ordinary radio pulsars \citep[$\sim$ 10$^{12}$ -- 10$^{13}$~G;
  see][for a review]{2013IJMPD..2230024T}.
Magnetic field decay and instabilities are recognized to be the engine
of the magnetar activity, characterized by both persistent and
bursting emission \citep{1995MNRAS.275..255T}. The former is ascribed
to thermal emission from the hot star surface, reprocessed by resonant
cyclotron scattering onto the charged particles in a twisted
magnetosphere with a luminosity $L_X$ $\sim$ 10$^{31}$ --
10$^{36}$~\lum. The latter consists of bursts and flares on different
time scales, ranging from few milliseconds to hundreds of seconds and
reaching luminosities up to 10$^{47}$~\lum. These bursting events are
often accompanied by an increase of the persistent flux up to three
orders of magnitude, which then usually relaxes back to the quiescent
level over months/years. The outbursts are most likely driven by
magnetic stresses, which result in elastic movements of the NS crust
and/or rearrangements/twistings of the external magnetic field
\citep{1995MNRAS.275..255T, 2011ApJ...727L..51P, 2011ApJ...741..123P},
with the formation of current-carrying localized bundles
\citep{2009ApJ...703.1044B,2012ApJ...750L...6P,2016ApJ...833..189L}.\\


\srclong\ (\src\ hereafter) was proposed as a magnetar because of the
detection of $\sim$ 10.6~s pulsations and the hot blackbody spectrum
\citep[\kt\ $\approx$ 0.6~keV;][]{2006ApJ...636L..41M,
  2006ApJ...653..587S}. An interesting feature is its association with
the young, massive Galactic star cluster Westerlund I. This provides
information about the NS progenitor: because of the young age of the
cluster ($\sim$ 4~Myr), the magnetar was likely produced by a star
with an initial mass $\gtrsim$ 40~M$_{\odot}$\footnote{To allow such a massive star to produce a NS, \citep{2014A&A...565A..90C} suggested a close binary comprising two stars of comparable masses ($\sim$ 41~M$_{\odot}$ + 35~M$_{\odot}$).}.

The magnetar nature of \src\ was confirmed when a short ($\sim$ 20~ms) and intense ($\sim$ 10$^{39}$~\lum\ in the 15 -- 150~keV energy band) burst triggered the {\em Neil Gehrels Swift Observatory} (\swift) Burst Alert Telescope (BAT) on 2006 September 21 \citep{2006ATel..894....1K}. About 12~hr later, the \swift\ X-ray Telescope (XRT) found the source in an enhanced flux state, $\sim$ 300 times brighter than four days earlier, when the source was at its historical minimum level (1 -- 10~keV absorbed flux of $\sim$ 1.5 $\times$ 10$^{-13}$~\flux; \citealt{2006ATel..902....1M}). A new outburst phase began five years later: on 2011 September 19, BAT detected few short bursts from a direction consistent with that of the source \citep{2011GCN.12359....1B} and a follow-up XRT observation showed a flux increase of a factor of $\sim$ 250 with respect to the pre-outburst level measured in 2006 September \citep*{2011ATel.3653....1I}. The spectral and timing properties of the 2006 outburst have been widely studied by several authors \citep{2007ApJ...664..448I, 2011ApJ...726...37W,2013ApJ...763...82A}.
\citet{2014MNRAS.441.1305R} presented an extended phase-coherent long-term timing solution and a phase-resolved analysis for both outbursts, using \cxo, \xmm\ and \swift\ data. They noted a similar evolution of the pulse profile in the two events: from a single-peaked structure during the quiescent state to a multi-peaked configuration in outburst.

The source entered a new bursting phase on 2017 May 16 when BAT
observed a burst from a location compatible with
\src\ \citep{GCN21095}. The XRT started to observe the field $\sim$
60~s after the trigger and the flux level was $\sim$ 10 times higher
than the quiescent level reached after the 2006 outburst (0.3 --
10~keV absorbed flux of $\sim$ 8 $\times$ 10$^{-13}$~\flux;
\citealt{2018MNRAS.474..961C}). We triggered our pre-approved
target-of-opportunity simultaneous observations with \cxo\ and \nus,
and started a \swift\ monitoring campaign to supplement these
pointings in order to study the evolution of the source spectral and
timing properties during the outburst decay. While recovering from
this last outburst, the source emitted two bursts that triggered BAT
on 2017 October 19 and 2018 February 5 \citep{2017ATel10877....1Y,
  2018ATel11264....1B}, producing two additional flux increases, the last
one being the larger of these three recent events. On the same days,
also the \fermi\ Gamma-Ray Burst Monitor detected bright and short
($\sim$ 0.1~s) SGR-like bursts from the source
\citep{2017GCN.22027....1R, GCN22402}.
 Moreover, \INT\ was triggered by two short bursts from a localization compatible with that of the magnetar on 2018 February 6 \citep{atel}. After this latest event, an \INT\ pointing was requested to study the soft gamma-ray emission that might have been associated with the bursts. The observation, however, did not detect any emission at the source position.


In this paper, we report on the results of \cxo, \nus\ and \swift\ observational campaigns covering the first $\sim$ 350 days of the outburst activity of \src\ after its re-activation in 2017 May. The analysis of the \INT\ pointings is also included. We first describe the data analysis procedure in Section \ref{sect:data}, then present the timing and spectral results in Section \ref{sect:timing} and \ref{sect:spectral}, respectively. Finally, we discuss our findings in Section \ref{sect:disc}.



\section{Observations and data reduction}
\label{sect:data}

Throughout this work we adopt the coordinates reported by
\citet{2006ApJ...636L..41M}, i.e. RA = 16$^h$47$^m$10$^{s}$.20, Dec =
-45$^{\circ}$52$^{\prime}$16$^{\prime\prime}$.9 (J2000.0), to convert
the photon arrival times to the Solar system barycentre reference
frame and the Solar system ephemeris DE200. A distance of 3.9~kpc is
assumed \citep{2007A&A...468..993K}. In the following, uncertainties
are quoted at 1$\sigma$ confidence level for a single parameter of
interest, otherwise noted. A log of the observations analysed in this paper is reported
in Tables \ref{tab:burst-spectra} and \ref{tab:log}.


\subsection{\swift}

\begin{figure*}
\centering
\resizebox{\hsize}{!}{\includegraphics[angle=0]{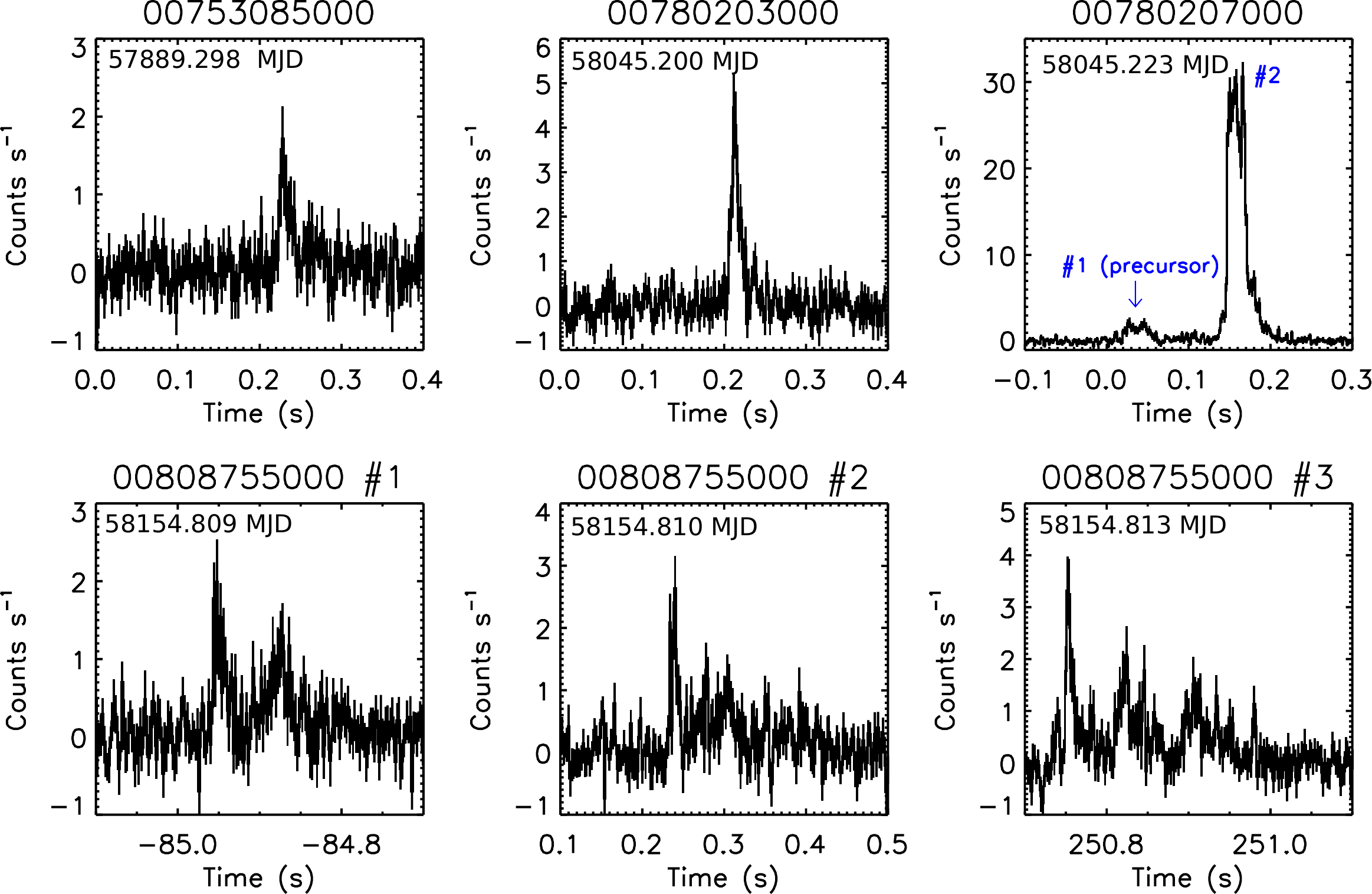}}
\caption{\label{fig:plotlcs} Light curves of the bright bursts detected from \srclong\ with \swift\ BAT (as indicated in each panel). The energy range is 15 -- 150~keV, the bin size 2~ms and the time is counted from the trigger time of each data set.}
\end{figure*}

For the observations where the BAT was triggered by bursts from \src\ (see Table \ref{tab:burst-spectra}), we created mask-tagged light curves from the event-mode data. The inspection of the light curves revealed the occurrence of one burst each in observations 00753085000 and 00780203000; in observation 00780207000, the powerful burst that alerted BAT followed a $\sim$ 0.13~s weaker event (a `precursor'), while in observation 00808755000, three bursts were recorded within a few minutes. To confirm that the emission was indeed associated with \src, for each of these events we verified the presence of a point source at the position of the magnetar in the 15 -- 150~keV sky images extracted across the burst duration (as computed by the Bayesian blocks algorithm \textsc{battblocks}). The same time intervals were used to extract the average spectra of the bursts. See Table \ref{tab:burst-spectra} for the time and duration of the bursts, and Figure \ref{fig:plotlcs} for their light curves.

From the outburst onset on 2017 May 16 until 2018 April 30, \src\ was
observed by XRT 47 times. The XRT was operating in photon counting
mode (PC; time resolution $\sim$ 2.51~s) and windowed timing mode
\citep[WT; time resolution $\sim$
  1.77~ms,][]{2005SSRv..120..165B}. The single exposure times ranged
from $\sim$ 0.5~ks to $\sim$ 5.5~ks. The monitoring campaign was
rather intense until the source entered a non-visibility window in
2017 October, just after the occurrence of the second burst. The
observations resumed in 2018 mid-January. Because of the flux
enhancement registered at the epoch of the third burst (2018 February
5), we asked to perform the subsequent observations in WT mode, to
mitigate possible pile-up effects. However, the flux rapidly decayed
over few days. During observations 00030806067 and 00030806068 (on
2018 March 2 and 10) \src\ was below the background level, therefore
we do not include these data sets in our analysis. The remaining
observations were hence carried out in PC mode.

We reprocessed the data and created exposure maps with {\sc xrtpipeline} (version 0.13.4, part of the {\sc heasoft} software package version 6.22) using the standard cleaning criteria. We selected events with grades 0 -- 12 and 0 for PC and WT mode, respectively\footnote{See \url{http://www.swift.ac.uk/analysis/xrt/digest_cal.php}.}. We extracted source counts from a circle with radius of 15 pixels centred on the source position (one XRT pixel corresponds to about 2.36\arcsec) for both PC and WT mode. Regarding the background estimation, we adopted an annulus with inner and outer radii of 40 and 80 pixels for the PC observations centred on the source position, while for WT data a region of the same size as that used for the source.
We applied the barycentre correction via the {\sc barycorr} tool. The spectra were generated by means of {\sc xselect} and the corresponding ancillary response files with the {\sc xrtmkarf} tool. We used the spectral redistribution matrices version `20130101v014' and `20131212v015' available in the calibration files for PC and WT data, respectively. In order to improve the source signal-to-noise ratio (S/N), we merged observations acquired within few days, after checking that no significant variability was present.

\subsection{\cxo}

\cxo\ observed \src\ five times between 2017 May and 2018 April, for a total dead-time corrected exposure time of $\sim$ 75.6~ks. All observations were performed with the Advanced CCD Imaging Spectrometer (ACIS-S; \citealt{2003SPIE.4851...28G}), set in timed exposure (TE) mode and with faint telemetry format (see Table \ref{tab:log} for a log). The source was always positioned on the back-illuminated S3 chip and 1/8 sub-array was adopted, resulting in a time resolution of $\sim$ 0.44~s. The data were processed following the standard analysis threads\footnote{\url{http://cxc.harvard.edu/ciao/threads/}} with the Chandra Interactive Analysis of Observations software ({\sc ciao}, version 4.9) and the calibration files {\sc caldb} 4.7.6.

Source and background photons were collected from a circular region with a radius of 2\arcsec\ and an annular region
with inner and outer radii of 4\arcsec\ and 10\arcsec, centred on the source position.
Photon arrival times were converted to the Solar system barycentre using the {\sc axbary} tool. Source and background spectra with the corresponding response matrices and ancillary files were created with {\sc specextract}.

\subsection{\nus}

\src\ was observed four times with \nus\ \citep{2013ApJ...770..103H}; these observations were coordinated with \cxo\ in order to probe the magnetar emission over a broader energy range thanks to \nus\ sensitivity to hard X-rays (3 -- 79~keV). The two focal plane modules FPMA and FPMB observed the source for a total dead-time corrected exposure time of $\sim$ 91.6~ks and 91.7~ks, respectively.

The data were reprocessed using the script {\sc nupipeline} of the \nus\ Data Analysis Software package ({\sc nustardas}, version 1.8.0) and the calibration files {\sc caldb} 20171002. We referred the event arrival times to the Solar system reference frame via the tool {\sc barycorr} and adopting the version 79 of the \nus\ clock file. Ghost ray contamination\footnote{Ghost rays are produced by photons reflected only once by the focusing mirrors. The source responsible is most likely the low mass X-ray binary GX\,340+0, situated at $\sim$ 20\arcmin\ from the magnetar.} is evident in the field of view for all the observations, affecting the detection of the magnetar. The source is detected in the 3 -- 8~keV energy band with a S/N of $\sim$ 14. The S/N does not increase when considering a broader energy band, suggesting that the source becomes background dominated above $\sim$ 8~keV. A circle with a 20\arcsec\ radius was used to collect source photons, while background counts were extracted from an annular region with radii of 70\arcsec\ and 130\arcsec, centred on the source. Using the {\sc nuproducts} tool, we produced light curves, background-subtracted spectra, instrumental response and auxiliary files for each FPM.

\subsection{\INT}

 The automatic \INT\ Bust Alert System \citep[IBAS,][]{ibas} detected two short bursts
in the IBIS/ISGRI data coming from a position compatible with
that of \src\ (note that the IBAS localisation accuracy is
3~arcmin at 90\% confidence level). The two events occurred on 
2018 February 6 at 07:25:56 UT (trigger ID 8007) and at 12:33:34 UT (trigger ID 8008). 
An offline search of this data set revealed the presence of a new burst at 00:33:19 UT. 

In order to search for additional bursts, we analysed all the available \INT\ data collected around the time of the aforementioned detections where 
the source was serendipitously observed within the field of view of the IBIS/ISGRI instrument \citep{ubertini03,lebrun03}. 
The data were reduced using version 10.2 of the Off-line Scientific Analysis software   
(OSA) distributed by the ISDC \citep{courvoisier03}. The \INT\ observations are divided into science windows (SCWs), 
i.e., pointings with typical durations of $\sim$ 2 -- 3~ks. We included in our analysis all SCWs where 
the source was located within 15 deg from the satellite aim direction, in order to minimise the instrument calibration 
uncertainties\footnote{\url{http://www.isdc.unige.ch/integral/analysis}}. The data set comprised SCWs in satellite revolutions 
1915 and 1916, spanning the time intervals from 2018 February 3 at 14:02 UT to February 4 at 
12:14 UT, and from 2018 February 5 at 21:09 UT to February 06 at 12:13 UT. The total exposure time was of 
141 ks. We also included all SCWs collected during a dedicated target-of-opportunity
observation performed in the direction of the source from 2018 February 
7 at 15:04 UT to February 11 at 23:45 UT (total exposure time of 170~ks; satellite revolution 1918). 
No bursts were found in all these data.

We noticed that the source was observed also at the rim of the IBIS/ISGRI field of view 
(off-axis angles between 15 and 17~degrees) between 2018 February 5 at 16:52 UT 
and at 21:09 UT (satellite revolution 1916). Although the instrument calibrations are slightly more uncertain at these 
higher off-axis angles, two strong bursts were clearly detected. For completeness, we mention that the typical
IBIS/ISGRI sensitivity to typical magnetar bursts during these large off-axis observations strongly
depends on time, with a median value of 1.7 $\times$ 10$^{-8}$~\flux\ for an 
integration time scale of 100~ms in the 25 -- 80~keV energy range.

\begin{table*}
\centering
\caption{Log of \swift\ BAT triggers of \srclong\ between 2017 May and 2018 February.}
\label{tab:burst-spectra}
\begin{tabular}{@{}lccc}
\hline
Burst$^a$ & UTC peak time & S/N$^b$ & $T_{90}$\,/\,total duration$^c$ \\
 & (YYYY-MM-DD hh:mm:ss) & & (s)  \\
\hline
00753085000 & 2017-05-16 07:09:02.127 & 8.4 & $0.018\pm0.004$ / 0.021  \\
00780203000 & 2017-10-19 04:48:48.193 & 13.7 & $0.016\pm0.004$ / 0.019 \\
00780207000 \#1 & 2017-10-19 05:20:39.695 & 13.9 & $0.031\pm0.006$ / 0.035 \\
00780207000 \#2 & 2017-10-19 05:20:39.826 & 55.1 & $0.034\pm0.003$ / 0.060  \\
00808755000 \#1 & 2018-02-05 19:25:46.830 & 11.1 & $0.106\pm0.018$ / 0.115  \\
00808755000 \#2 & 2018-02-05 19:27:11.968  & 8.5 & $0.008\pm0.002$ / 0.009  \\
00808755000 \#3 & 2018-02-05 19:31:22.582 & 19.1 & $0.184\pm0.021$ / 0.206  \\
\hline
\end{tabular}
\begin{list}{}{}
\item[$^{a}$] The notation \#N indicates corresponds to the burst number in a given observation.
\item[$^{b}$] Source signal-to-noise ratio in the 15 -- 150\,keV image.
\item[$^{c}$] The $T_{90}$ duration is the time during which 90\% of the burst counts were accumulated.  The total duration is computed by the Bayesian blocks algorithm {\sc BATTBLOCKS}.
\end{list}
\end{table*}

\begin{table*}
\caption{Log of the X-ray observations of \srclong\ between 2017 May 16th and 2018 April 28th.}
\label{tab:log}
\footnotesize
\resizebox{2.1\columnwidth}{!}{
\begin{tabular}{@{}lcccccc}
\hline
Obs. ID         &  Instrument$^*$  & Mid date & Start time (TT) & End time (TT)     & Exposure            & source net count rate$^{**}$   \\
                &              &  (MJD)    & \multicolumn{2}{c}{(YYYY-MM-DD hh:mm:ss)}      & (ks) & (counts s$^{-1}$) \\
\hline
00753085000 & \swift/XRT & 57889.303 & 2017/05/16 07:10:18 & 2017/05/16 07:21:08 & 0.6 & 0.104 $\pm$ 0.013 \\
00030806033    & \swift/XRT & 57892.970 & 2017/05/19 02:28:46 & 2017/05/20 20:04:53 & 4.7 & 0.065 $\pm$ 0.004 \\
19135          & \cxo/ACIS-S  & 57898.074 & 2017/05/25 00:09:57 & 2017/05/25 03:22:23 & 9.1 & 0.223 $\pm$ 0.005 \\
80201050002    & \nus/FPMA & 57900.298 & 2017/05/27 01:46:09 & 2017/05/27 12:31:09 & 15.7 & 0.012 $\pm$ 0.001 \\
80201050002    & \nus/FPMB & 57900.298 & 2017/05/27 01:46:09 & 2017/05/27 12:31:09 & 15.5 &  0.009 $\pm$ 0.001 \\
00030806034    & \swift/XRT & 57907.235 & 2017/06/03 03:50:06 & 2017/06/03 07:25:54 & 4.7 & 0.046 $\pm$ 0.003 \\
00030806035    & \swift/XRT & 57910.159 & 2017/06/06 00:38:59 & 2017/06/06 06:57:53 & 3.7 & 0.052 $\pm$ 0.004 \\
00030806036    & \swift/XRT & 57913.349 & 2017/06/09 06:31:59 & 2017/06/09 10:11:54 & 5.1 & 0.065 $\pm$ 0.004 \\
19136          & \cxo/ACIS-S & 57920.141 & 2017/06/16 01:05:02 & 2017/06/16 05:42:02 & 13.7 & 0.237 $\pm$ 0.004 \\
80201050004    & \nus/FPMA  & 57921.483 & 2017/06/17 05:11:09 & 2017/06/17 18:01:09 & 21.7  & 0.017 $\pm$ 0.001 \\
80201050004    & \nus/FPMB  & 57921.483 & 2017/06/17 05:11:09 & 2017/06/17 18:01:09 & 21.6  & 0.014 $\pm$ 0.001 \\
00030806037    & \swift/XRT & 57922.662 & 2017/06/18 10:55:52 & 2017/06/18 20:49:54 & 3.9  & 0.055 $\pm$ 0.004 \\
00030806038    & \swift/XRT & 57934.406 & 2017/06/30 09:39:21 & 2017/06/30 09:48:52 & 0.5  & 0.048 $\pm$ 0.010 \\
00030806039    & \swift/XRT & 57937.196 & 2017/07/03 01:24:26 & 2017/07/03 08:00:52 & 3.9  & 0.052 $\pm$ 0.004 \\
00030806040    & \swift/XRT & 57943.955 & 2017/07/09 22:02:26 & 2017/07/09 23:49:53 & 1.2  & 0.047 $\pm$ 0.006 \\
19137          & \cxo/ACIS-S & 57944.403 & 2017/07/10 06:37:30 & 2017/07/10 12:43:59 & 18.2 & 0.228 $\pm$ 0.004  \\
80201050006    & \nus/FPMA  & 57948.582 & 2017/07/14 07:51:09 & 2017/07/14 20:06:09 & 22.3 & 0.015 $\pm$ 0.001 \\
80201050006    & \nus/FPMB  & 57948.582 & 2017/07/14 07:51:09 & 2017/07/14 20:06:09 & 22.8 & 0.019 $\pm$ 0.001 \\
00030806041    & \swift/XRT & 57949.561 & 2017/07/15 10:20:14 & 2017/07/15 16:34:52 & 2.3  &  0.056 $\pm$ 0.005 \\
00030806042    & \swift/XRT & 57951.496 & 2017/07/17 02:17:45 & 2017/07/17 21:29:52 & 3.8 & 0.058 $\pm$ 0.004 \\
00030806043    & \swift/XRT & 57953.801 & 2017/07/19 14:27:14 & 2017/07/19 23:59:52 & 1.5 & 0.051 $\pm$ 0.006\\
00030806044    & \swift/XRT & 57958.322 & 2017/07/24 01:20:41 & 2017/07/24 14:05:52 & 3.7 & 0.054 $\pm$ 0.004 \\
00030806045    & \swift/XRT & 57965.468 & 2017/07/31 03:48:06 & 2017/07/31 18:38:52 & 2.9 & 0.055 $\pm$ 0.004 \\
00030806046$^d$    & \swift/XRT & 57969.159 & 2017/08/04 03:46:20 & 2017/08/04 03:51:54 & 0.3 & 0.052 $\pm$ 0.013 \\
00030806047$^d$    & \swift/XRT & 57974.646 & 2017/08/09 14:43:54 & 2017/08/09 16:15:52 & 0.7 & 0.057 $\pm$ 0.009 \\
00030806048    & \swift/XRT & 57978.785 & 2017/08/13 15:32:41 & 2017/08/13 22:07:53 & 3.3 & 0.047 $\pm$ 0.004 \\
00030806049    & \swift/XRT & 57981.686 & 2017/08/16 14:01:52 & 2017/08/16 18:53:52 & 1.5 & 0.057 $\pm$ 0.006 \\
00030806050    & \swift/XRT & 57993.437 & 2017/08/27 21:09:16 & 2017/08/28 23:49:52 & 0.9 & 0.052 $\pm$ 0.008 \\
00030806051    & \swift/XRT & 58006.584 & 2017/09/10 06:35:31 & 2017/09/10 21:26:52 & 5.4 & 0.050 $\pm$ 0.003 \\
00030806052    & \swift/XRT & 58020.448 & 2017/09/24 01:05:51 & 2017/09/24 20:23:53 & 1.6 & 0.056 $\pm$ 0.006 \\
00030806053    & \swift/XRT & 58023.567 & 2017/09/27 11:51:57 & 2017/09/27 15:19:52 & 3.1 & 0.052 $\pm$ 0.004 \\
00030806054    & \swift/XRT & 58033.556 & 2017/10/07 04:27:26 & 2017/10/07 22:12:53 & 4.6 & 0.043 $\pm$ 0.003 \\
00030806055    & \swift/XRT & 58038.273 & 2017/10/12 00:55:33 & 2017/10/12 12:10:51 & 4.5 & 0.050 $\pm$ 0.003 \\
00780203000    & \swift/XRT & 58045.383 & 2017/10/19 04:50:42 & 2017/10/19 13:31:13 & 13.1 & 0.078 $\pm$ 0.002 \\
00030806056    & \swift/XRT & 58046.709 & 2017/10/20 06:33:41 & 2017/10/21 03:27:52 & 4.5 & 0.066 $\pm$ 0.004 \\
00030806057    & \swift/XRT & 58138.199 & 2018/01/19 23:57:40 & 2018/01/20 09:36:52 & 3.0 & 0.045 $\pm$ 0.004 \\
00030806058$^e$    & \swift/XRT & 58139.335 & 2018/01/21 07:53:53 & 2018/01/21 08:09:53 & 0.9 & 0.038 $\pm$ 0.006 \\
00030806059$^e$    & \swift/XRT & 58141.919 & 2018/01/23 21:58:42 & 2018/01/23 22:06:53 & 0.5 & 0.048 $\pm$ 0.010 \\
00030806060    & \swift/XRT & 58143.526 & 2018/01/25 02:59:01 & 2018/01/25 22:15:53 & 2.7 & 0.042 $\pm$ 0.004 \\
00030806061    & \swift/XRT & 58144.887 & 2018/01/26 20:22:43 & 2018/01/26 22:12:52 & 1.9 & 0.038 $\pm$ 0.005 \\
00030806062    & \swift/XRT & 58146.181 & 2018/01/28 01:04:38 & 2018/01/28 07:36:52 & 4.9 & 0.038 $\pm$ 0.003 \\
00808755000    & \swift/XRT & 58154.818 & 2018/02/05 19:28:19 & 2018/02/05 19:48:21 & 1.2 & 0.284 $\pm$ 0.016\\
00030806064    & \swift/XRT (WT) & 58156.371 & 2018/02/07 03:23:29 & 2018/02/07 14:25:56 & 2.9 & 0.138 $\pm$ 0.008 \\
00030806065    & \swift/XRT & 58160.688 & 2018/02/11 01:13:41 & 2018/02/12 07:49:51 & 5.2 & 0.068 $\pm$ 0.004 \\
19138$^f$          & \cxo/ACIS-S & 58174.053 & 2018/02/24 22:09:30 & 2018/02/25 04:24:51 & 18.2 & 0.286 $\pm$ 0.004 \\
20976$^f$          & \cxo/ACIS-S & 58174.748 & 2018/02/25 15:09:14 & 2018/02/25 20:46:11 & 16.4 & 0.278 $\pm$ 0.004 \\
00030806066    & \swift/XRT & 58174.863 & 2018/02/25 17:27:40 & 2018/02/25 23:59:53 & 4.9 & 0.069 $\pm$ 0.004 \\
80201050008    & \nus/FPMA & 58176.276 & 2018/02/26 19:31:09 & 2018/02/27 17:46:09 & 31.9 & 0.013 $\pm$ 0.001 \\
80201050008    & \nus/FPMB & 58176.276 & 2018/02/26 19:31:09 & 2018/02/27 17:46:09 & 31.8 & 0.014 $\pm$ 0.001 \\
00030806069    & \swift/XRT & 58194.637 & 2018/03/17 06:34:43 & 2018/03/17 23:59:54 & 2.9 & 0.042 $\pm$ 0.003 \\
00030806070    & \swift/XRT & 58201.805 & 2018/03/24 16:43:54 & 2018/03/24 21:54:53 & 4.3 & 0.058 $\pm$ 0.004 \\
00030806071    & \swift/XRT & 58209.450 & 2018/04/01 08:21:20 & 2018/04/01 13:14:52 & 1.1 & 0.057 $\pm$ 0.007 \\
00030806072    & \swift/XRT & 58215.524 & 2018/04/07 10:59:09 & 2018/04/07 14:08:51 & 1.7 & 0.046 $\pm$ 0.005 \\
00030806073$^g$ & \swift/XRT & 58219.596 & 2018/04/11 05:24:09 & 2018/04/11 23:13:10 & 0.4 & 0.064 $\pm$ 0.013 \\
00030806074$^g$ & \swift/XRT & 58220.956 & 2018/04/12 22:52:14 & 2018/04/12 23:03:53 & 0.7 & 0.066 $\pm$ 0.009 \\
00030806075    & \swift/XRT & 58222.190 & 2018/04/14 03:29:34 & 2018/04/14 05:37:54 & 2.1 & 0.062 $\pm$ 0.005 \\
00030806076    & \swift/XRT & 58229.605 & 2018/04/21 12:46:50 & 2018/04/21 16:15:54 & 2.2 & 0.066 $\pm$ 0.005 \\
00030806077    & \swift/XRT & 58236.834 & 2018/04/28 18:18:24 & 2018/04/28 21:45:54 & 2.7 & 0.044 $\pm$ 0.004 \\
\hline
\hline
\end{tabular}
}
\begin{list}{}{}
\item[$^*$] \swift\ XRT operated in PC mode, otherwise specified. \cxo\ ACIS-S was set in TE mode.
\item[$^{**}$] For \cxo\ and \swift\ XRT-PC observations the source net count rate refers to the 0.3 -- 10~keV energy band, while for XRT-WT ones to the 1 -- 10~keV range. For \nus\ it corresponds to the 3 -- 8 keV energy interval.
\item[$^d$,$^e$,$^f$,$^g$] These observations were merged in the spectral analysis.
\end{list}
\end{table*}


\begin{figure}
\begin{center}
\includegraphics[scale=0.31]{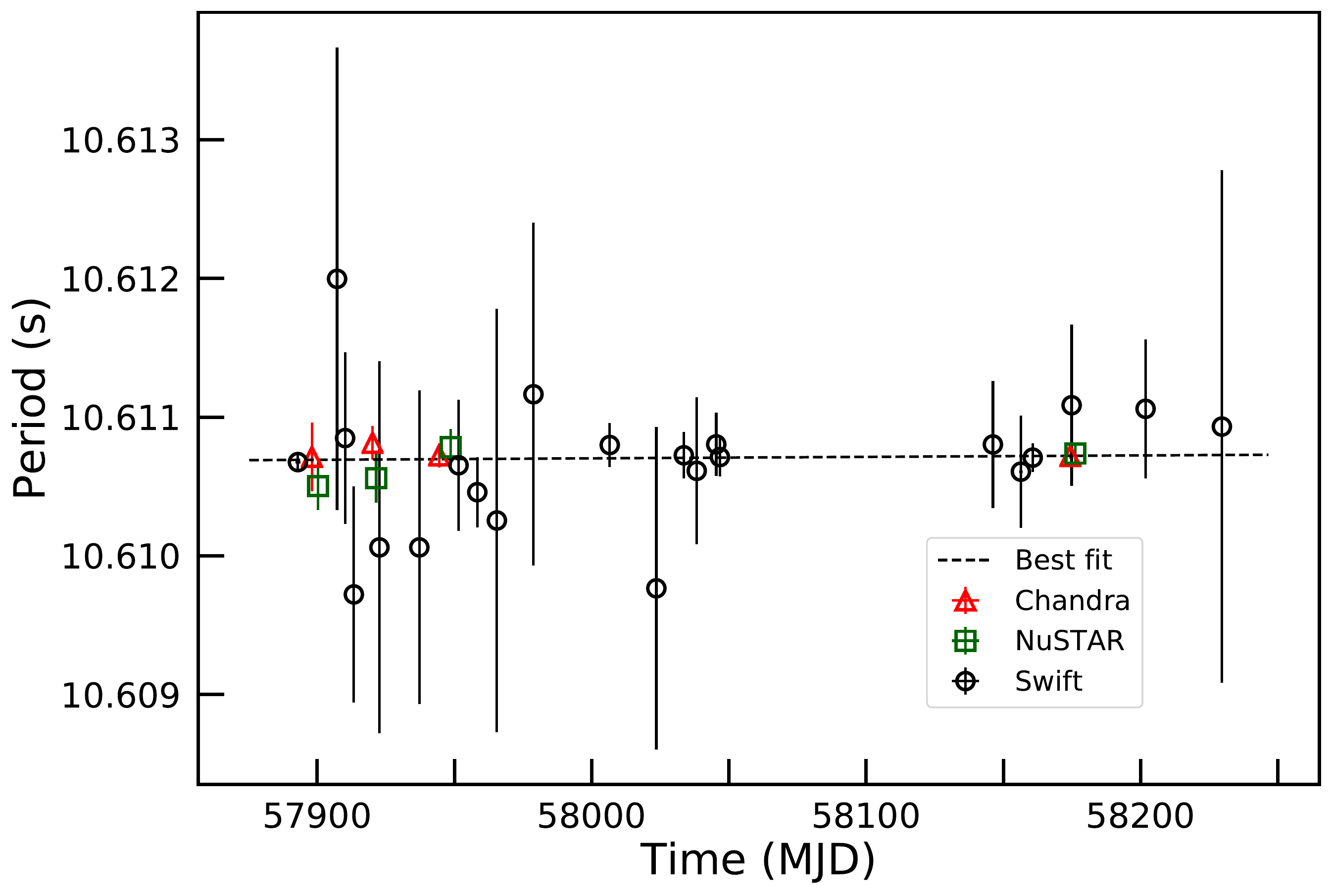}
\end{center}
\caption{Temporal evolution of the spin period after the \swift\ BAT trigger on 2017 May 16. Red triangles, green squares and  black circles are the measurements from \cxo, \nus\ and \swift\ observations, respectively. The dashed line is the linear function which fits the data best (see text for details).}

\label{fig:timing}
\vskip -0.1truecm
\end{figure}

\section{Timing analysis}
\label{sect:timing}

Timing studies for the previous (2006 and 2011) outbursts have been performed by several authors, applying both phase-coherent and non-coherent techniques \citep{2007ApJ...664..448I, 2011ApJ...726...37W, 2013ApJ...763...82A, 2014MNRAS.441.1305R}. In the past, the source exhibited a pulse profile that changed during the outbursts, showing the transition from a simpler morphology to a multi-peaked structure.

For our analysis, we selected events in the 0.3 -- 8~keV energy band for \cxo, 0.3 -- 10~keV for \swift\ and 3 -- 8~keV for \nus. For the latter, we combined the FPMA and FPMB event files for each observation. First, we tried to build a phase-coherent timing solution starting from the \cxo\ observation 19138, which was performed about 20 days after the last burst. The Fourier spectrum showed a prominent peak at the spin frequency of \src, $\sim$ 10.6~s, and strong harmonic content up to the second harmonic (confirming the pulse profile complex structure close to a bursting activity period). We applied a phase-fitting technique to extend the solution over a longer time interval, but we could not find a solution that aligned all the profiles. We note that a phase-coherent analysis requires to be able to track unambiguously the phase evolution with time of a reference structure in the pulse profile. Due to the different time resolutions, \cxo\ and \nus\ pulse profiles showed two peaks, while in most \swift\ profiles the distinction between the two peaks was not clear, making the choice of a reference structure more complicated. 


%

Therefore, we decided to use a different approach to constrain the average spin down rate. We searched for the spin period in each observation by means of the $Z^2_n$ test \citep{1983A&A...128..245B}. Given the approximate knowledge of the source period, we run the test in the 10.60 -- 10.62~s period range, with the number $n$ of harmonics fixed to 2. We performed Monte Carlo simulations to determine the uncertainty of the best period \citep[for details see][]{1999ApJ...522L..49G}.
We then fit the best periods as a function of time with a linear function, $P(t) = P_0 + \dot{P} t$. The best-fitting parameters were $P_0$ = 10.608(3)~s and $\dot{P}$ = (1 $\pm$ 2) $\times$ 10$^{-12}$~\ss. The period derivative we measured is consistent with zero, but this does not imply that the source is not spinning down. The data used for the timing analysis do not provide enough sensitivity to measure $\dot{P}$ values as small as those previously obtained for this source. Therefore, we derive the 90\% upper limit, 4 $\times$ 10$^{-12}$~\ss. We note that the obtained upper limit is higher than the estimates reported in previous works \citep[see Table 2 by][]{2014MNRAS.441.1305R}. Figure \ref{fig:timing} shows the time evolution of the spin period and the best-fitting linear model.\\

\begin{figure*}
\begin{center}
\includegraphics[scale=0.4]{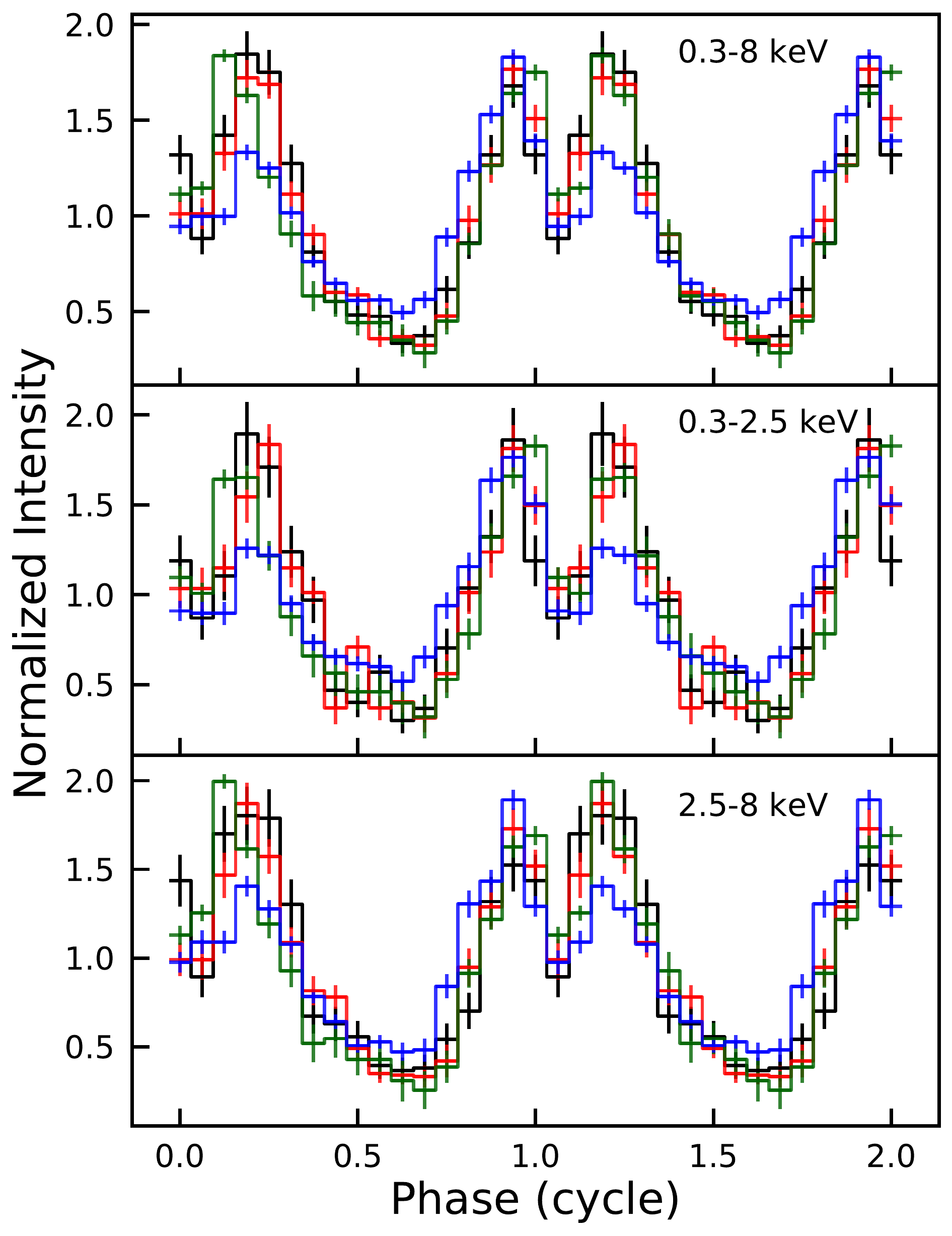}
\includegraphics[scale=0.4]{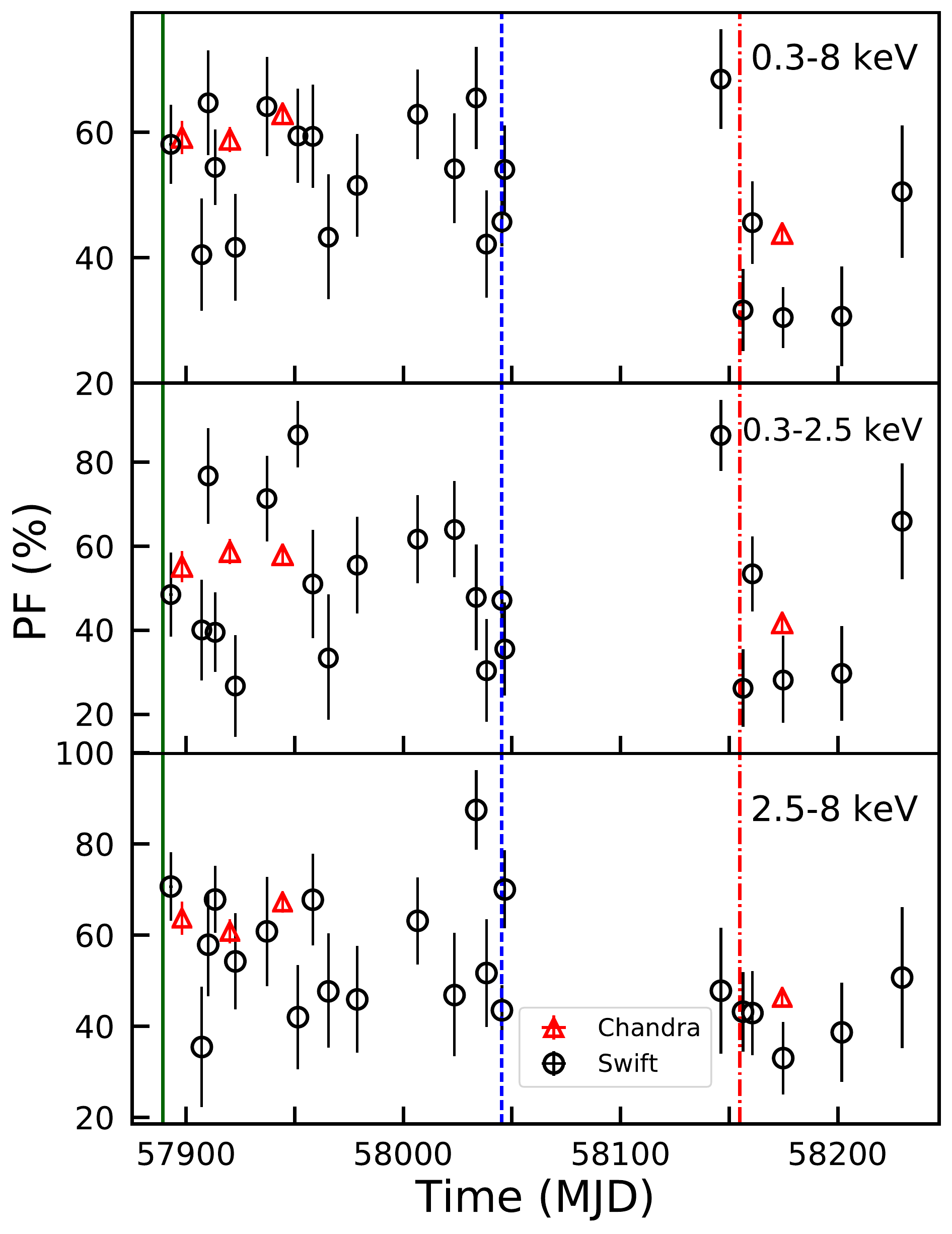}
\end{center}
\caption{ {\it Left}: Pulse profiles of \srclong\ obtained from \cxo\ observations (black: ID 19135, red: ID 19136, green: ID 19137, blue: merged event file for IDs 19138 and 20976) in different energy bands: 0.3 -- 8~keV (top panel); 0.3 -- 2.5~keV (middle panel); 2.5 -- 8~keV (bottom panel). The pulse profiles are shifted along the horizontal axis for phase alignment and sampled in 16 phase bins. Two cycles are shown for better visualization. {\it Right}: Pulsed fraction as a function of time for \cxo\ (red triangles) and \swift\ (black circles) pointings in different energy bands: 0.3 -- 8~keV (top panel); 0.3 -- 2.5~keV (middle panel); 2.5 -- 8~keV (bottom panel). The vertical lines denote the epochs of the three BAT triggers: 2017 May 16 at 07:09:02 UT (solid green line), 2017 October 19 at 04:48:48 UT (dashed blue line) and 2018 February 05 at 19:27:11 UT (dash-dotted red line).   }

\label{fig:pulseprofile}
\vskip -0.1truecm
\end{figure*}

Next, we folded the \cxo\ and \swift\ background-subtracted and
exposure-corrected light curves on the best period determined in each
observation. We studied the shape of the pulse profiles in different
energy bands: 0.3 -- 8~keV, 0.3 -- 2.5~keV (soft band) and 2.5 --
8~keV (hard bard). We chose these energy intervals so as to have
comparable photon counting statistics. The \cxo\ pulse profiles
presented a multi-peaked configuration, well modelled by a combination
of three sinusoidal functions plus a constant (see Figure
\ref{fig:pulseprofile}, left panel), while the \swift\ profiles could
only be reproduced by a constant plus one sine, given the lower
statistics and the fact that the time resolution of the PC mode
($\sim$ 2.5~s) is unable to sample accurately the complex profile
structure.

%

Furthermore, we computed the pulsed fraction (defined as the semi-amplitude of the fundamental divided by the average count rate) in the same energy bands and studied its temporal evolution (see Figure \ref{fig:pulseprofile}, right panel). We noted that in all the three energy bands, the pulsed fraction dropped after the last burst and in the last observation it seemed to recover the average pre-outburst value, $\sim$ 48\% for the total band, $\sim$ 52\% for the soft band and $\sim$ 60\% for the hard band.


%
%
%

\section{Spectral analysis}
\label{sect:spectral}

The spectral analysis was performed with the {\sc xspec} package (version 12.9.1m; \citealt{1996ASPC..101...17A}). Once the best fit was found, the absorbed and unabsorbed fluxes were estimated with the convolution model {\sc cflux}. For the luminosity quiescent level, we adopted the value 2.6 $\times$ 10$^{33}$~\lum, derived by \citet{2013MNRAS.434..123V} with a resonant Compton scattering (RCS) model from the \xmm\ observation performed on 2006 September 16.

\subsection{The BAT burst events}


We fit all the burst spectra in the 15 -- 150~keV energy range with single-component models typically used for magnetar bursts, such as a power-law (PL), a blackbody (BB) and a bremsstrahlung (BREMSS) component. 


The three models provided a statistically equivalent description of the spectra relative to the observations 00753085000, 00780203000, the second and third burst detected in the trigger 00808755000. In Table \ref{tab:burst-fitting} we report the results relative to the blackbody model. For the first burst in observation 00808755000 and the `precursor' in observation 00780207000, the blackbody model did not give an acceptable fit. The best-fitting values for a power-law model are listed in Table \ref{tab:burst-fitting}. The inclusion of an additional component, in terms of another blackbody, was required for the main event in observation 00780207000 ($F$-test probability of $\sim$ 3 $\times$ 10$^{-12}$ for a two-blackbody model). 


The most powerful event was the burst that triggered BAT on 2017 October 19 at 05:20:52 UT (trigger 780207); it reached a luminosity of $\sim$ 9 $\times$ 10$^{39}$~\lum\ in the 15 -- 150~keV energy band and the `precursor' was about one order of magnitude weaker. The event, which occurred $\sim$ 30~min before (at 04:48:48 UT, trigger 780203), was as intense as the precursor, $L$ $\sim$ 1.5 $\times$ 10$^{39}$~\lum; the other bursts have a luminosity in the range (5 -- 9) $\times$ 10$^{38}$~\lum.

\begin{table*}
\centering
\caption{Spectral analysis results for the bursts from \srclong\ detected by \swift\ BAT. }
\resizebox{2.1\columnwidth}{!}{
\label{tab:burst-fitting}
\begin{tabular}{@{}lccccccc}
\hline
Burst$^a$ & Model & $kT_1$\,/\,$R_1$ & $kT_2$\,/\,$R_2$ & $\Gamma$ & Flux$^b$ & Fluence & $\chi_\nu^2$ (dof)$^c$ \\
 &  & (keV)\,/\,(km) & (keV)\,/\,(km) & & (10$^{-7}$~\flux) & (erg~cm$^{-2}$) & \\
\hline
00753085000 & BB &  4.2 $\pm$ 0.8 / 0.5$^{+0.4}_{-0.1}$ &  &  & 2.5 $\pm$ 0.4 & (5.4 $\pm$ 0.8) $\times$ 10$^{-9}$  & 1.4 (17) \\
00780203000 & BB &  7.1 $\pm$ 0.6 / 0.3 $\pm$ 0.1 &  &  & 9.2 $\pm$ 0.8 & (1.7 $\pm$ 0.1) $\times$ 10$^{-8}$ & 1.5 (21)\\
00780207000 \#1 & PL &  &  & 2.3 $\pm$ 0.2 & 7.0 $\pm$ 0.6 & (2.4 $\pm$ 0.2) $\times$ 10$^{-8}$ & 1.0 (28)\\
00780207000 \#2 & 2BB & 5.1 $\pm$ 0.5 / 0.9 $\pm$ 0.2 & 12.4 $\pm$ 0.8 / 0.14 $\pm$ 0.02 & & 49.7 $\pm$ 1.1 & (2.98 $\pm$ 0.06) $\times$ 10$^{-7}$ & 0.7 (35) \\
00808755000 \#1 & PL &  &  & 1.8 $\pm$ 0.2 & 3.3 $\pm$ 0.4 & (3.7 $\pm$ 0.4) $\times$ 10$^{-8}$ & 0.6 (28) \\
00808755000 \#2 & BB & 9.1 $\pm$ 1.6 / 0.2 $\pm$ 0.1 &  &  & 7.5 $\pm$ 1.2 & (6.7 $\pm$ 1.1) $\times$ 10$^{-9}$ & 0.8 (16) \\
00808755000 \#3 & BB & 10.2 $\pm$ 0.4 / 0.03 $\pm$ 0.01 &  &  & 2.8 $\pm$ 0.2 & (5.8 $\pm$ 0.3) $\times$ 10$^{-9}$ & 1.5 (36) \\
\hline
\end{tabular}
}
\begin{list}{}{}
\item[$^{a}$] The notation \#N indicates corresponds to the burst number in a given observation.
\item[$^{b}$] In the 15 -- 150~keV energy range.
\item[$^{c}$] $\chi_\nu^2$ is the reduced chi-squared and dof stands for `degrees of freedom'.
\end{list}
\end{table*}

\subsection{The \INT\ upper limits}

 For the observations where bursts were not detected, we estimated a typical sensitivity for IBIS/ISGRI 
to the typical burst emission at 5$\sigma$ confidence level at 
7.9 $\times$ 10$^{-9}$~\flux, considering an integration time scale of 100~ms in the energy range 25 -- 80~keV. 

\begin{figure*}
    \centering
    \includegraphics[scale=0.25]{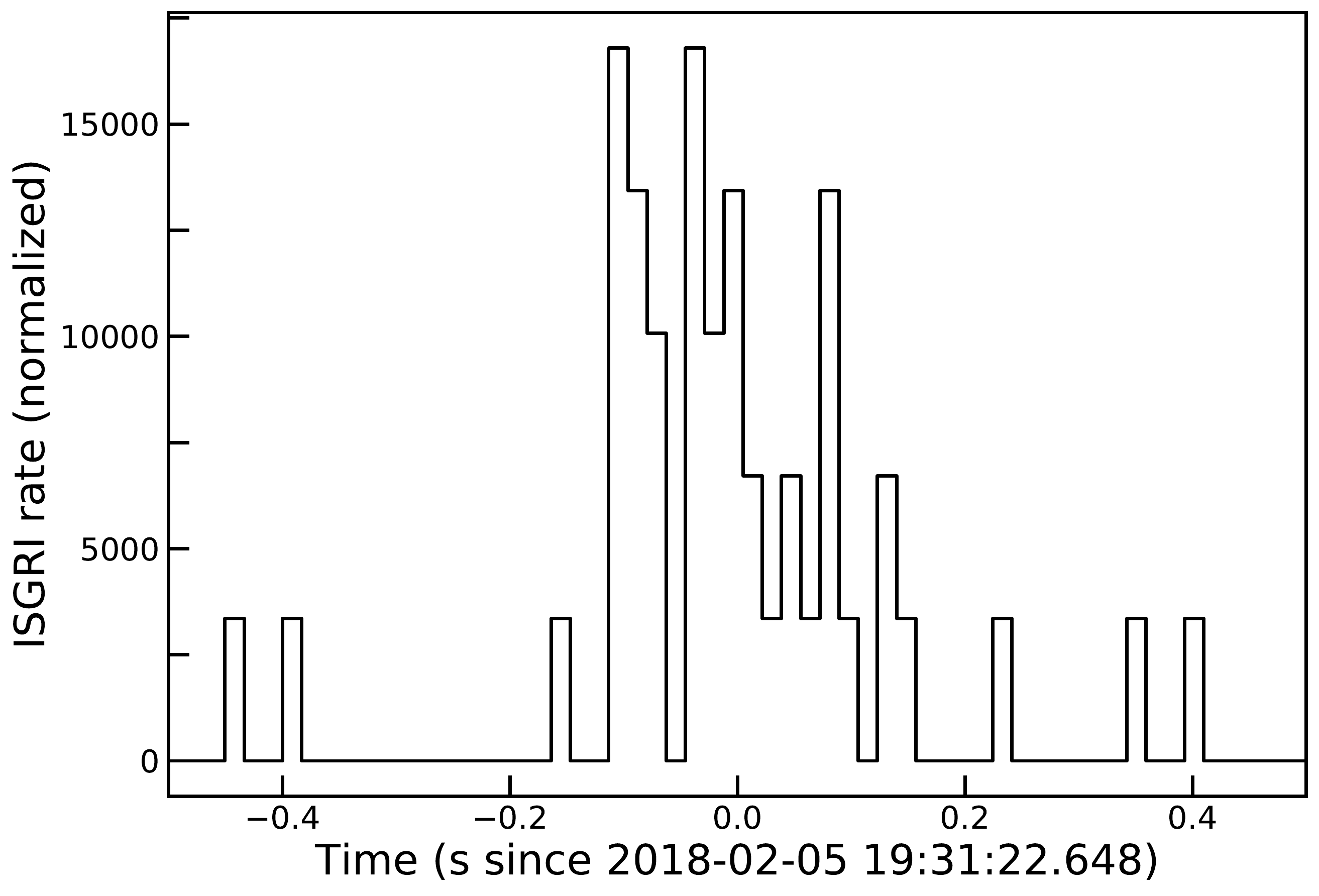} 
    \includegraphics[scale=0.25]{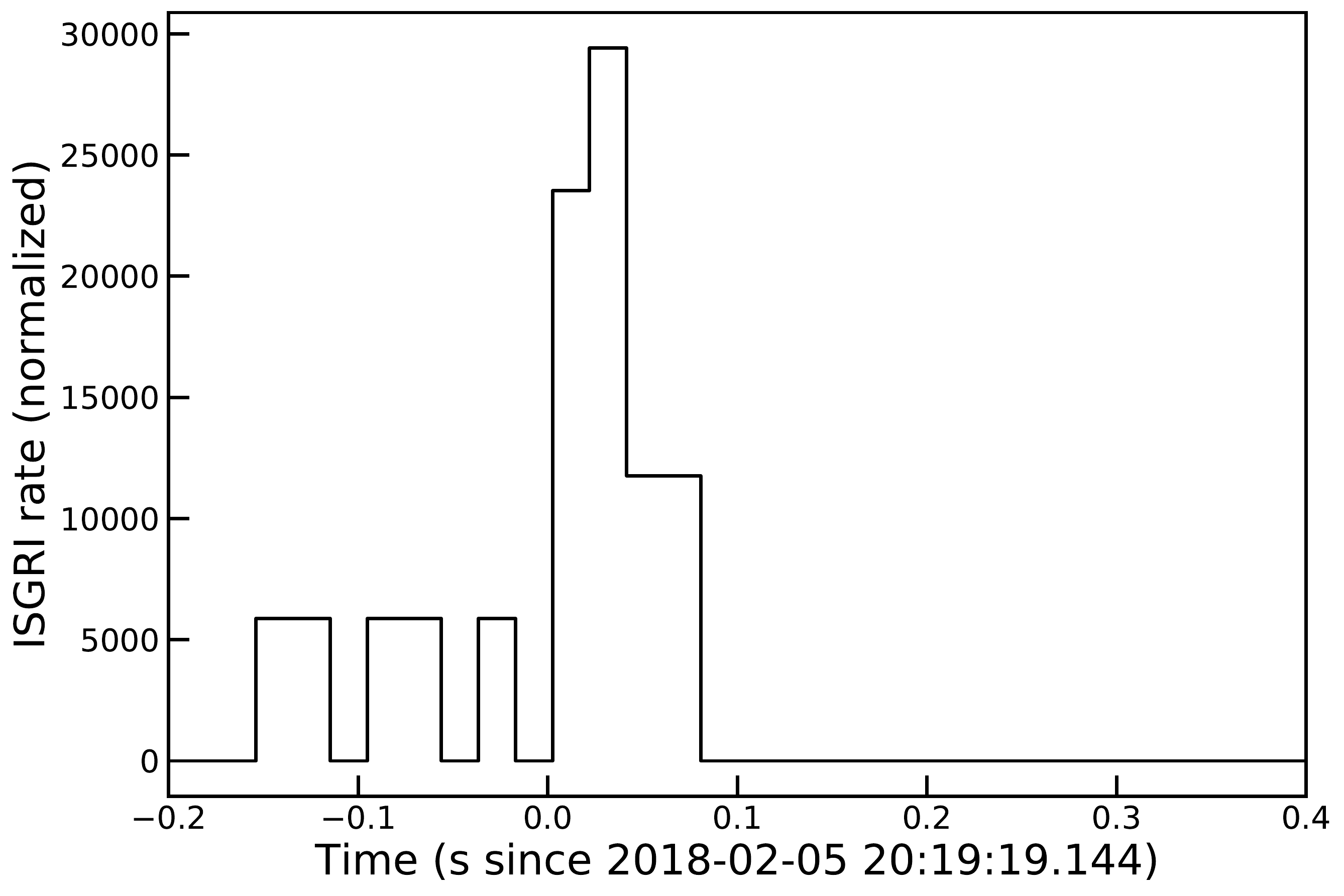} 
    \includegraphics[scale=0.25]{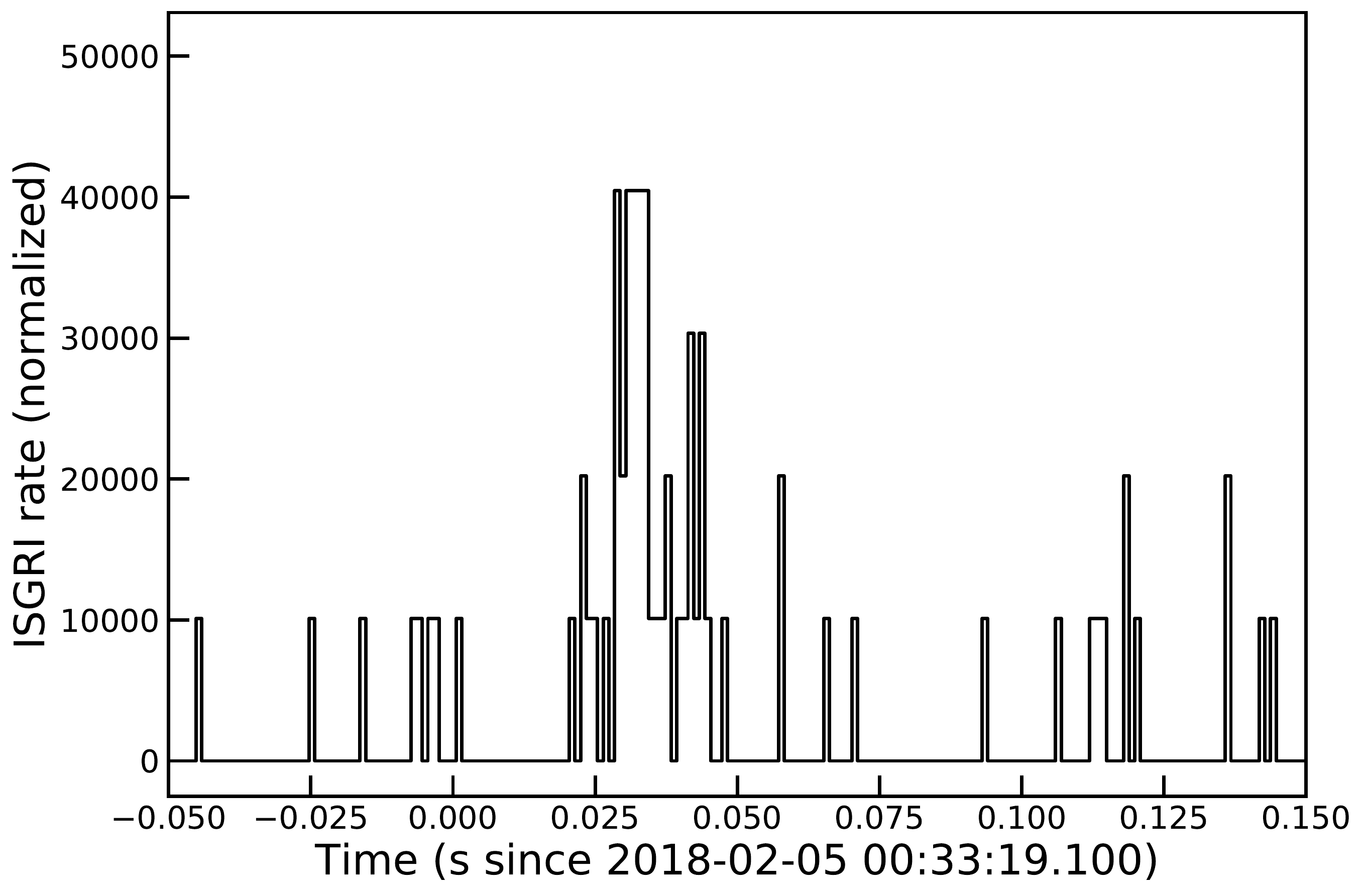} 
    \includegraphics[scale=0.25]{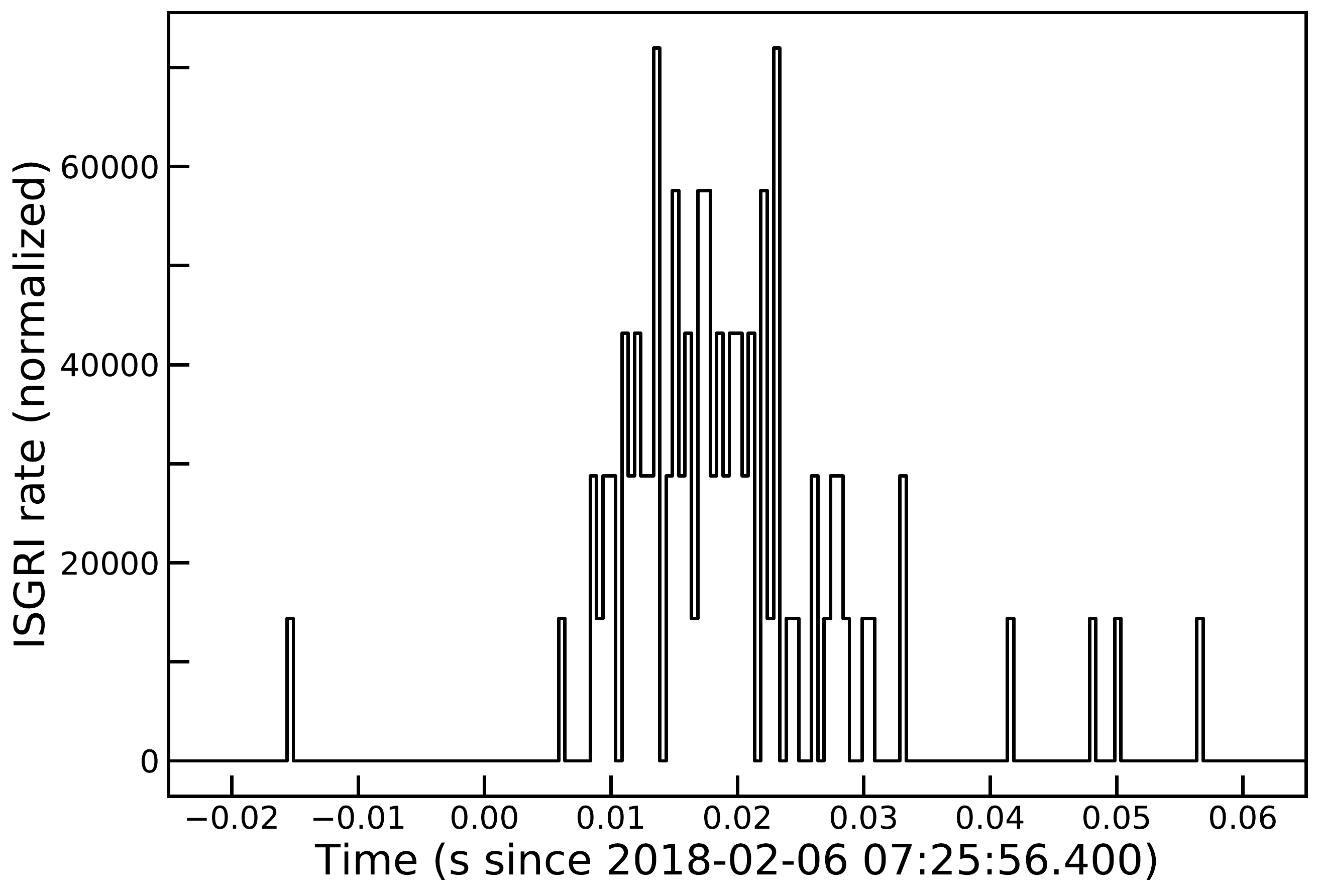} 
    \includegraphics[scale=0.25]{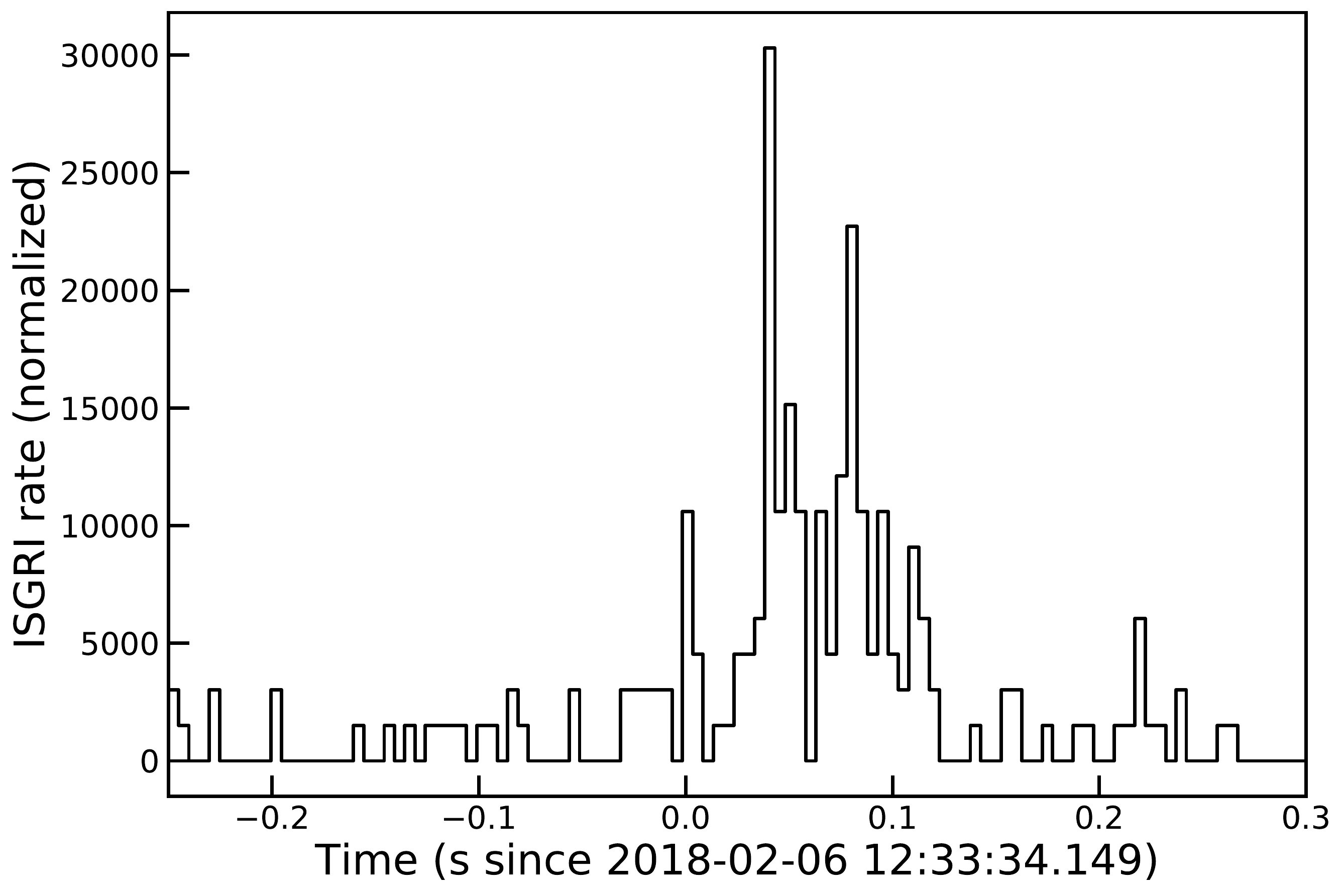} \\
    \caption{IBIS/ISGRI light curves of all detected bursts during the satellite revolutions 1915-1918. 
    The displayed count rates are corrected for vignetting and refer to the 20 -- 80~keV energy band.}
    \label{fig:my_label}
\end{figure*}

The two bursts observed by \INT\ on 2018 February 5 were also independently 
detected by \swift\ BAT and \fermi\ Gamma-Ray Burst Monitor. We report the times and fluences of all bursts detected by IBIS/ISGRI in Table~\ref{tab:int_bursts}, and show the corresponding light curves in Figure~\ref{fig:my_label}.


No persistent emission from the source could be detected by IBIS/ISGRI in any of the individual or combined 
SCWs in revolutions 1915-1918. By stacking all the data together, we obtained an upper limit on the source 
persistent emission of 3.5 $\times$ 10$^{-11}$~\flux\ 
in the 20 -- 80~keV energy range at 5$\sigma$ confidence level (total effective exposure time of 474.1~ks). 

\begin{table}
    \centering
    \caption{Times and fluences in the 25 -- 80~keV energy range of the bursts from \srclong\
    detected by the IBIS/ISGRI on-board \INT\ during the satellite revolutions 1915-1918.}
    \label{tab:int_bursts}
    \begin{tabular}{@{}cc}
\hline    
Trigger time (UTC)	& Fluence \\
(YYYY-MM-DD hh:mm:ss) & (10$^{-8}$ erg~cm$^{-2}$) \\
\hline
      2018-02-05 19:31:22 &       7.1~$\pm$~0.8 \\
      2018-02-05 20:19:19 &       7.3~$\pm$~0.9 \\
      2018-02-06 00:33:19 &       1.7~$\pm$~0.4 \\
      2018-02-06 07:25:56 &       2.0~$\pm$~0.5 \\
      2018-02-06 12:33:34 &       2.5~$\pm$~0.6 \\  
\hline
 
\end{tabular}
\end{table}

\subsection{The X-ray monitoring}
\label{sect:monitoring}

Because of the different effective areas of the X-ray instruments that translate into different counting statistics, we preferred to fit the \swift\ XRT data separately from the \cxo\ and \nus\ ones. We adopted the model {\sc tbabs} to describe the photoelectric absorption along the line of sight with photoionization cross-sections from \citet{1996ApJ...465..487V} and chemical abundances from \citet*{2000ApJ...542..914W}.\\

First, we present the results of the \swift\ XRT monitoring campaign.
The \swift\ background-subtracted spectra were rebinned according to a
minimum number of counts variable from observation to observation. In
most cases, we used less than 10 counts per spectral bin, with the
exception of three observations (IDs: 00780203000, 00808755000 and
00030806064) where the larger number of counts was enough to adopt a
higher grouping minimum (at least 20 counts per bin). For the
\swift\ spectra, we applied the $W$ statistic (suited for Poisson
distributed data with Poisson distributed background\footnote{In {\sc
    Xspec} the $W$ statistic is turned on with the command {\tt
    statistic cstat} and if a background has been read. See
  \url{https://heasarc.gsfc.nasa.gov/docs/xanadu/xspec/wstat.ps} and
  \url{https://heasarc.gsfc.nasa.gov/xanadu/xspec/manual/XSappendixStatistics.html}.}). We
restricted our spectral modelling to the 0.3 -- 10~keV energy band for
the PC data, while for the WT mode spectra the energy channels below
$\sim$ 1~keV were ignored due to known calibration issues\footnote{See
  \url{http://www.swift.ac.uk/analysis/xrt/digest_cal.php}.}. As a
first step, we fit the spectra individually using an absorbed
blackbody model ({\sc tbabs*bbodyrad}). This model provided a good fit
to all the observations, except for observations 00780203000 and
00808755000, which are the XRT pointings following the BAT triggers
for the latest two bursting events. 

 We fit these two spectra simultaneously with an absorbed blackbody 
plus power law model (BB+PL hereafter), forcing the hydrogen column density to be
the same across the two data sets. The simultaneous fit yielded \nh\ = (3.4 $\pm$ 0.7) $\times$
10$^{22}$~\cm2. The fit for the observation 00780203000 gave the following parameters:
blackbody temperature \kt\ = 0.5 $\pm$ 0.1~keV and radius \r\ = 1.1$^{+2.0}_{-0.2}$~km plus a 
power law with photon index $\Gamma$ = 2.1$^{+0.6}_{-0.8}$. The other data set (ID: 00808755000)
is well described by a blackbody with \kt\ = 0.5 $\pm$ 0.1~keV and \r\ = 1.8$^{+1.7}_{-0.5}$~km 
and a power law with $\Gamma$ = 0.4$^{+0.9}_{-1.1}$.

\begin{figure*}
\begin{center}
\includegraphics[scale=0.52]{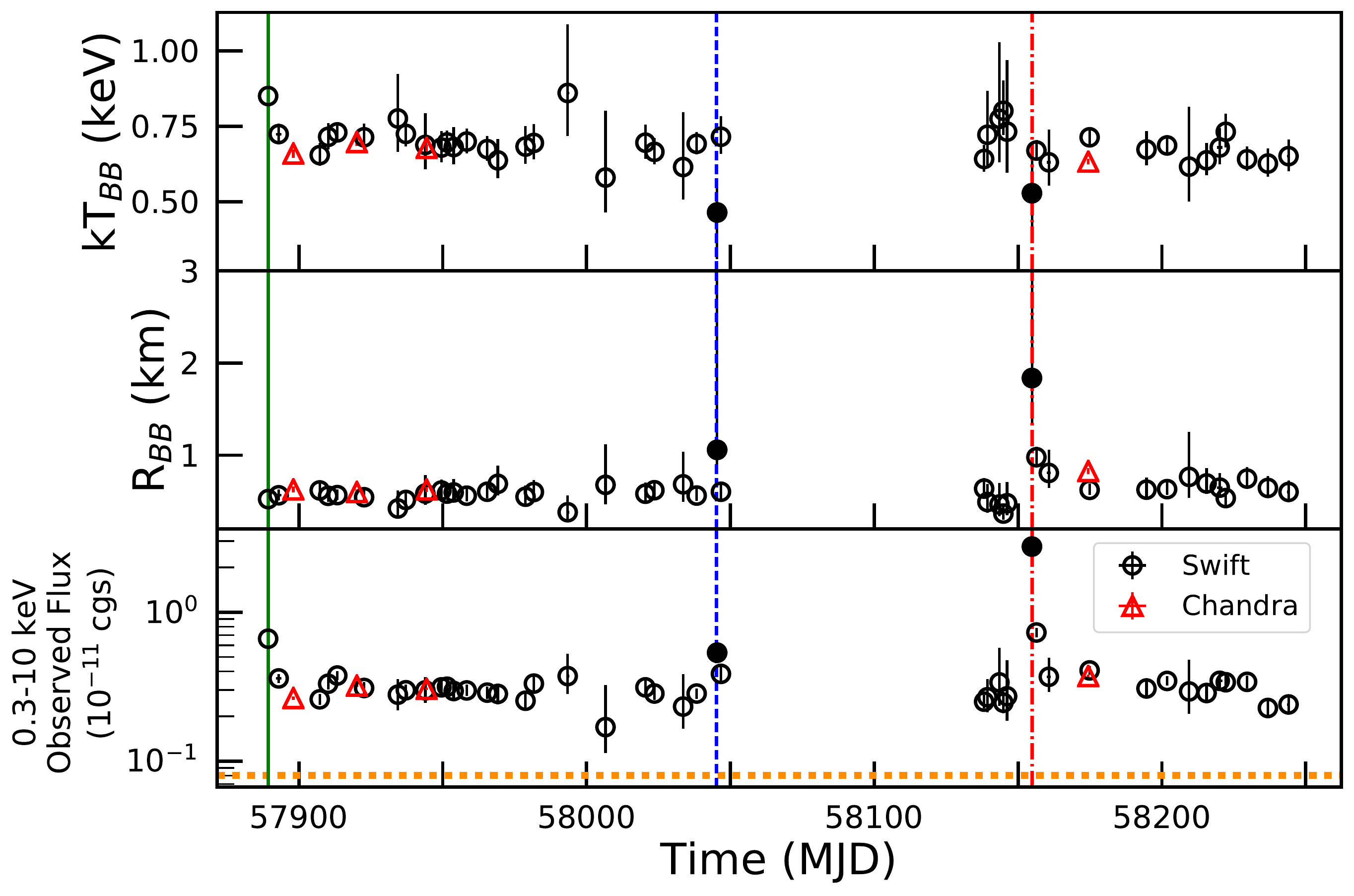}
\end{center}
\caption{Temporal evolution of the blackbody temperature (top
panel) and radius (middle panel) calculated at infinity,
assuming a 3.9~kpc distance. In the bottom panel,
temporal evolution of the absorbed flux in the 0.3 -- 10~keV
energy range; the dotted orange line indicates the flux level that
the source reached after the 2006 outburst, 8 $\times$
10$^{-13}$~\flux\ \citep{2018MNRAS.474..961C}. Red triangles are relative
to the \cxo\ pointings and black circles
represent \swift\ XRT observations; the filled circles denote the spectra where
a BB+PL model is required. The vertical lines denote the epochs of
the three BAT triggers: 2017 May 16 at 07:09:02 UT (solid green
line), 2017 October 19 at 04:48:48 UT (dashed blue line) and 2018
February 05 at 19:27:11 UT (dash-dotted red line).}

\label{fig:bb}
\vskip -0.1truecm
\end{figure*}

In addition to the individual modelling, we fit all the spectra together removing the two above-mentioned ones. The hydrogen column density was constrained to be the same across all the data sets, while the blackbody parameters were left free to vary. The value of \nh, inferred from the simultaneous fit, was (2.5 $\pm$ 0.1) $\times$ 10$^{22}$~\cm2; the temporal evolution of the blackbody temperature and radius is shown in Figure \ref{fig:bb}, top and middle panels. 

The quality of the fit was evaluated performing Monte Carlo simulations; we used the command {\sc goodness} in {\sc xspec} to simulate 1000 spectra whose parameters are drawn from Gaussian distributions centred on the best-fit values with width equal to the derived 1$\sigma$ uncertainty. The percentage of simulations with the test statistic less than that for the data ranged from 40\% to 60\%. Figure \ref{fig:spectra}, left panel, shows the spectra for the observations 00753085000 (May 2017), 00780203000 (Oct 2017) and 00808755000 (Feb 2018) with the respective best-fit models and residuals; these observations are the XRT re-pointings after the BAT triggers. In chronological order, the 0.3 -- 10~keV unabsorbed fluxes are (9 $\pm$ 1) $\times$ 10$^{-12}$, (1.5$^{+1.7}_{-0.4}$) $\times$ 10$^{-11}$ and (4.3$^{+0.8}_{-0.5}$ ) $\times$ 10$^{-11}$~\flux, which translate into a luminosity of (1.8 $\pm$ 0.2) $\times$ 10$^{34}$, (2.8 $\pm$ 0.8) $\times$ 10$^{34}$ and (7.8 $\pm$ 1.4) $\times$ 10$^{34}$~\lum. The Feb 2018 event marked the highest enhancement of the X-ray persistent flux among the registered bursting activities, reaching a luminosity a factor $\sim$ 30 higher than in the quiescent level.


The \cxo\ background-subtracted spectra were grouped using the optimal binning scheme of \citet{2016A&A...587A.151K} by means of the ftool {\sc ftgrouppha} and fitted in the 0.3 -- 8~keV energy range, using the \chisq\ statistic. We merged observations 19138 and 20976, being only one day apart, having similar count rates and because no significant spectral variability was found. We estimated the impact of pile-up with Web{\sc pimms} and found that its fraction ranges from 3.5\% to 4.5\% across the different observations\footnote{In Web{\sc pimms} the estimated pileup percentage is defined as the ratio of the number of frames with two or more events to the number of frames with one or more events times 100.}. To correct for this effect, we included the multiplicative pile-up model \citep{2001ApJ...562..575D}, as implemented in {\sc xspec}, in the spectral fitting procedure. Following `The \cxo\ ABC guide to Pileup'\footnote{See \url{http://cxc.harvard.edu/ciao/download/doc/pileup_abc.pdf}.}, we allowed the grade migration parameter $\alpha$ to vary and fixed the parameter {\em psffrac} equal to 0.95, i.e. we assumed that 95\% of events are within the central, piled-up portion of the source point spread function. The parameter $\alpha$ was forced to be the same across the different observations because of the similar count rates. We fit simultaneously the four spectra applying a blackbody corrected by the pile-up model and tying the hydrogen column up across the different observations. The fit yielded a \nh\ = (2.9 $\pm$ 0.1) $\times$ 10$^{22}$~\cm2 (\rchisq\ = 1.0 for 284 dof); the other spectral parameters, the fluxes and luminosities are reported in Table \ref{tab:results_figurespectra}. Figure \ref{fig:spectra}, right panel, shows the spectra with the best-fit model and the corresponding residuals.

The \nus\ spectra were rebinned with at least 20 counts per bin. Since the spectrum is background dominated over $\sim$ 8~keV, these data sets are insufficient to characterize properly the hard X-ray emission of \src, but can provide a further check for \cxo\ results. We fit the \nus\ spectra simultaneously with the \cxo\ ones acquired at the same epoch; the inclusion of these new observations did not affect the spectral analysis results. Moreover, we verified that the values of the spectral parameters did not show any dependence on the choice of the size for the background region.


\begin{table*}

\begin{minipage}{1.\textwidth}

\begin{center}
\includegraphics[scale=0.3]{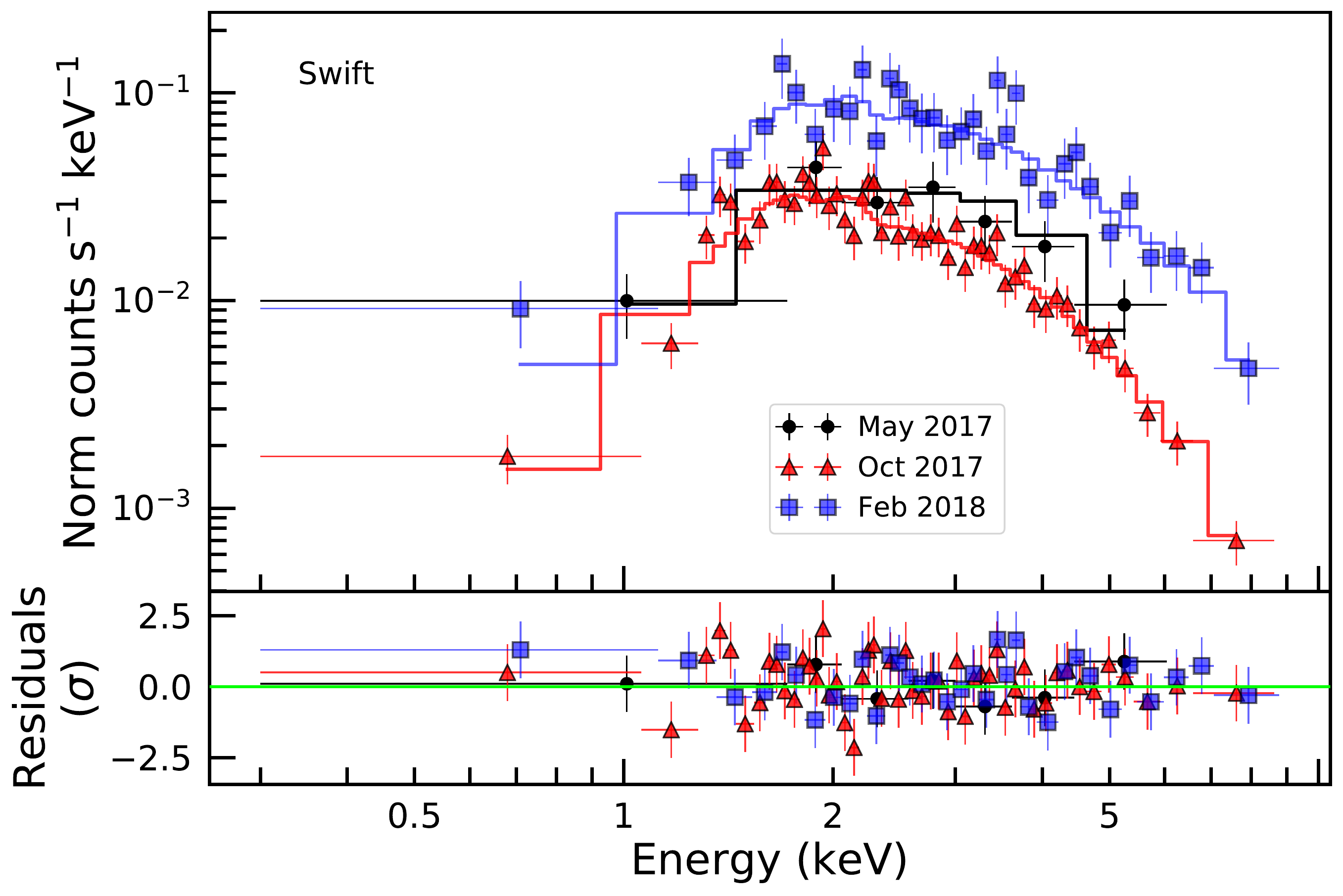}
\includegraphics[scale=0.3]{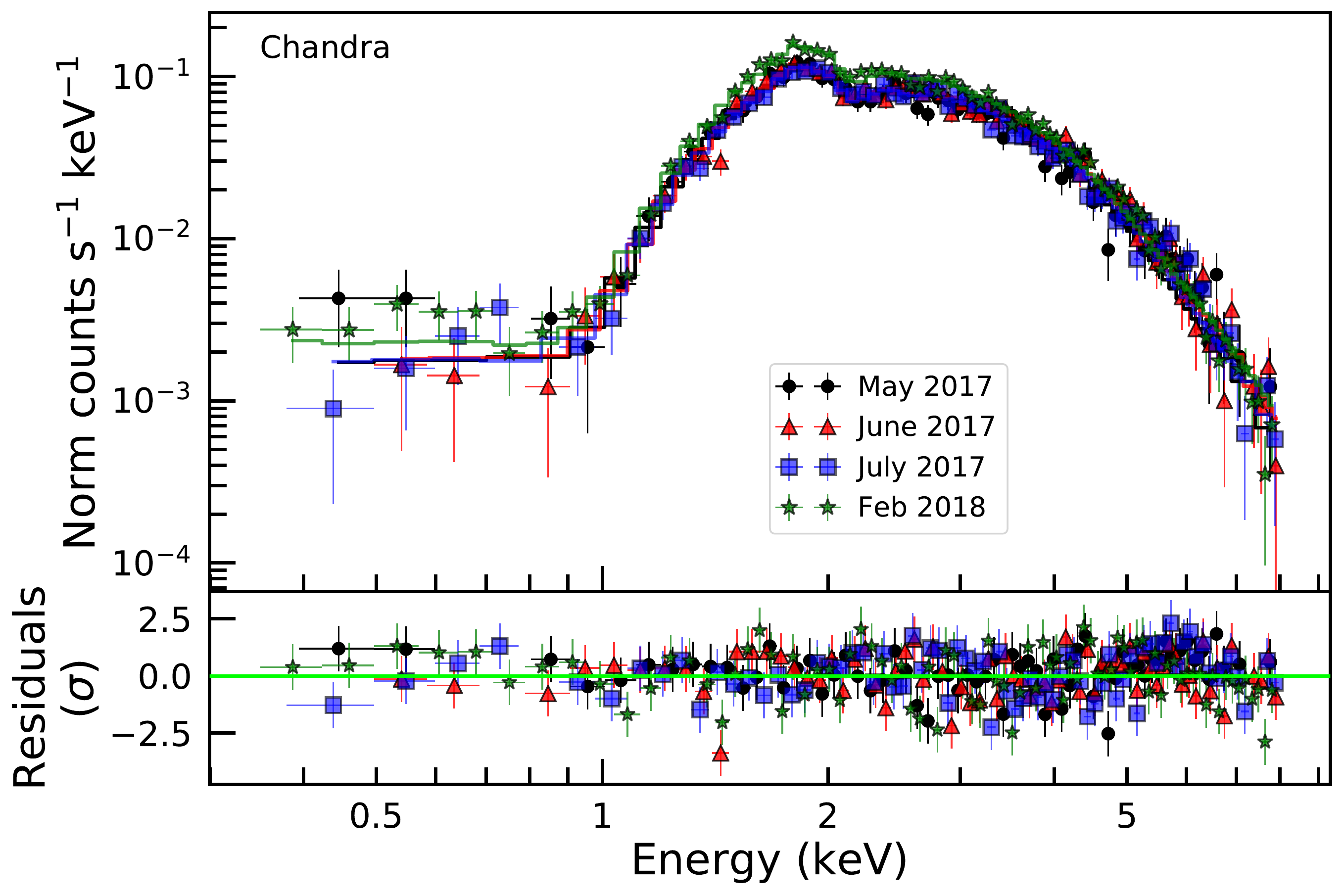}
\end{center}
\captionof{figure}{ {\it Left}: \swift\ XRT spectra corresponding to the observations following the BAT triggers with the best-fit models (solid line) and the  post-fit residuals in units of standard deviations. {\it Right}: \cxo\ spectra fitted with an absorbed blackbody corrected by the pileup model and the residuals with respect to this combined model. For more details see Section \ref{sect:monitoring}.}


\label{fig:spectra}
\vskip -0.1truecm

\end{minipage} 

\begin{minipage}{1.\textwidth}
\centering
\caption{Spectral analysis results for the \swift\ XRT (above the line) and \cxo\ spectra plotted in Figure \ref{fig:spectra}. }
\label{tab:results_figurespectra}
\begin{tabular}{@{}lcccccc}
\hline
ID & $kT$ & $R$ & $\Gamma$ & PL Norm$^a$ & Flux$_{abs}$ $^b$ & $L_X$ $^b$ \\
 &  (keV) & (km) &  & 10$^{-3}$ & (10$^{-12}$~\flux) & (10$^{34}$~\lum) \\
\hline
00753085000 & 0.8 $\pm$ 0.1 & 0.5 $\pm$ 0.1 & -- & -- & 6.6 $\pm$ 1.0 & 1.8 $\pm$ 0.2 \\  
00780203000 & 0.5 $\pm$ 0.1 & 1.1$^{+2.0}_{-0.2}$ & 2.1$^{+0.6}_{-0.8}$ & 2.3 $\pm$ 0.2 & 5.3 $\pm$ 0.3 & 2.8 $\pm$ 0.8 \\ 
00808755000 & 0.5 $\pm$ 0.1 & 1.8$^{+1.7}_{-0.5}$ & 0.4$^{+0.9}_{-1.1}$ & 0.6$^{+2.9}_{-0.5}$ & 27.6 $\pm$ 2.6 & 7.8 $\pm$ 1.4 \\
\hline
19135 & 0.66 $\pm$ 0.01 & 0.63 $\pm$ 0.03 & -- & -- & 2.6 $\pm$ 0.1 & 0.95 $\pm$ 0.02 \\
19136 & 0.70 $\pm$ 0.01 & 0.60 $\pm$ 0.03 & -- & -- & 3.21 $\pm$ 0.07 & 10.9 $\pm$ 0.2 \\
19137 & 0.68 $\pm$ 0.01 & 0.62 $\pm$ 0.03 & -- & -- & 3.05 $\pm$ 0.06 & 10.7 $\pm$ 0.2 \\
19138+20976 & 0.63 $\pm$ 0.01 & 0.83 $\pm$ 0.02 & -- & -- & 3.72 $\pm$ 0.04 & 14.0 $\pm$ 0.03 \\
\hline
\end{tabular}

\begin{list}{}{}
\item[$^{a}$] The power law normalization is in units of photons/keV/cm$^2$ at 1~keV.
\item[$^{b}$] In the 0.3 -- 10~keV energy range.
\end{list}
\end{minipage}

\end{table*}




\section{Discussion}
\label{sect:disc}


We have presented the evolution of the spectral and timing properties of the magnetar \src\ following its latest
outburst activity, which started with the detection of short X-ray bursts in 2017 May. Our monitoring campaign covered
$\sim$ 350 days of the outburst evolution, allowing us to characterise accurately the behaviour of the source over a
long time span. In the last observation, performed on 2018 April 28, the observed 0.3 -- 10~keV flux was
(2.4 $\pm$ 0.3) $\times$ 10$^{-12}$~\flux, about 15 times higher than the historical minimum measured by
\xmm\ in 2006, four days before the first known outburst activation \citep{2006ATel..902....1M}.\\

\src\ underwent three bursting episodes during this latest
activation, entering the small list of magnetars showing
recurrent outburst activity, including, e.g., 1E\,1048$-$59,
1E\,1547$-$5408, SGR\,1627$-$41 and 1E\,2259$+$586.
The emission of short bursts is accompanied by a considerable enhancement of the X-ray persistent
flux (see the flux evolution and the long-term light curve in Figure \ref{fig:bb}, bottom
panel, and Figure \ref{fig:lumlongterm}, respectively). To obtain a detailed description
of the temporal evolution of the 0.3 -- 10~keV luminosity, we
modelled the decay pattern following each episode separately using
a combination of a constant $L_q$ plus one or more exponential
functions, depending on the shape of the light curve:
\begin{equation}
L(t)= L_q + \sum_{i=1}^2 A_i\times \rm{exp}(-$$t$$/\tau_i)~,
\end{equation}
where the $e$-folding time $\tau_i$ can be considered as
an estimate of the decay time scale, similarly to the analysis performed by
\citet{2018MNRAS.474..961C}.

For the first two flux enhancements in 2017 May and October, 
the source did not reach the historical quiescent level before the onset of the following event. 
In these two cases, the constant $L_q$ was held fixed to the quiescent 
value attained after that particular event, i.e. 9.3 $\times$ 10$^{33}$~\lum\ 
and 8.4 $\times$ 10$^{33}$~\lum\ for 2017 May and October events, respectively.
The best-fitting model is represented by a simple exponential function in both 
cases with $e$-folding times $\tau_{May}$ = 2.4$_{-0.6}^{+1.0}$~d and $\tau_{Oct}$ = 1.3$_{-0.4}^{+1.1}$~d,
which reflect the fast decay at the early stage of these bursting events. 
For the last outburst episode in 2018 February, the constant was fixed to the
quiescent value 2.6 $\times$ 10$^{33}$~\lum\ \citep{2013MNRAS.434..123V}.  
In this case, a double-exponential function was required to properly fit the
decay with $e$-folding times $\tau_{1,Feb}$ =
0.8$_{-0.1}^{+0.3}$~d and $\tau_{2,Feb}$ = 167$_{-39}^{+73}$~d,
the latter tracking the long-term decay.
We computed the energy released in these outburst episodes by integrating the
best-fitting model for the time evolution of the luminosity over
the whole duration of the event. The onset of an event is determined by the corresponding BAT triggers. 
For the first two episodes, the end was set by the beginning of the following event; while  
for the last one, we extrapolated the epoch of recovery of the quiescent state.
During the 2017 May and October events, the source released an energy equal to $\sim$ 1.3 $\times$ 10$^{41}$~erg
and $\sim$ 8.2 $\times$ 10$^{40}$~erg, respectively. 
%
For the latest event, our decay fit predicts that the source
will return in quiescence around 2019 October, releasing a total
energy of $\sim$ 3.2 $\times$ 10$^{41}$~erg. This value is
estimated assuming no change in the decay pattern, and should
hence be considered only as a rough estimate. 

The case of \src\ is analogous to that of SGR\,1627$-$41 and
1E\,1547$-$5408, which did not fully recover from their first
outbursts in 1998 and 2008, respectively, before resuming a new
outburst activity. On the other hand, the case of
1E\,1048$-$59 is slightly different since the outbursts seem to
be periodic, and the source always returns to its
quiescent level before entering a new outburst episode \citep{2015ApJ...800...33A}.

\begin{figure}
\begin{center}
\includegraphics[scale=0.33]{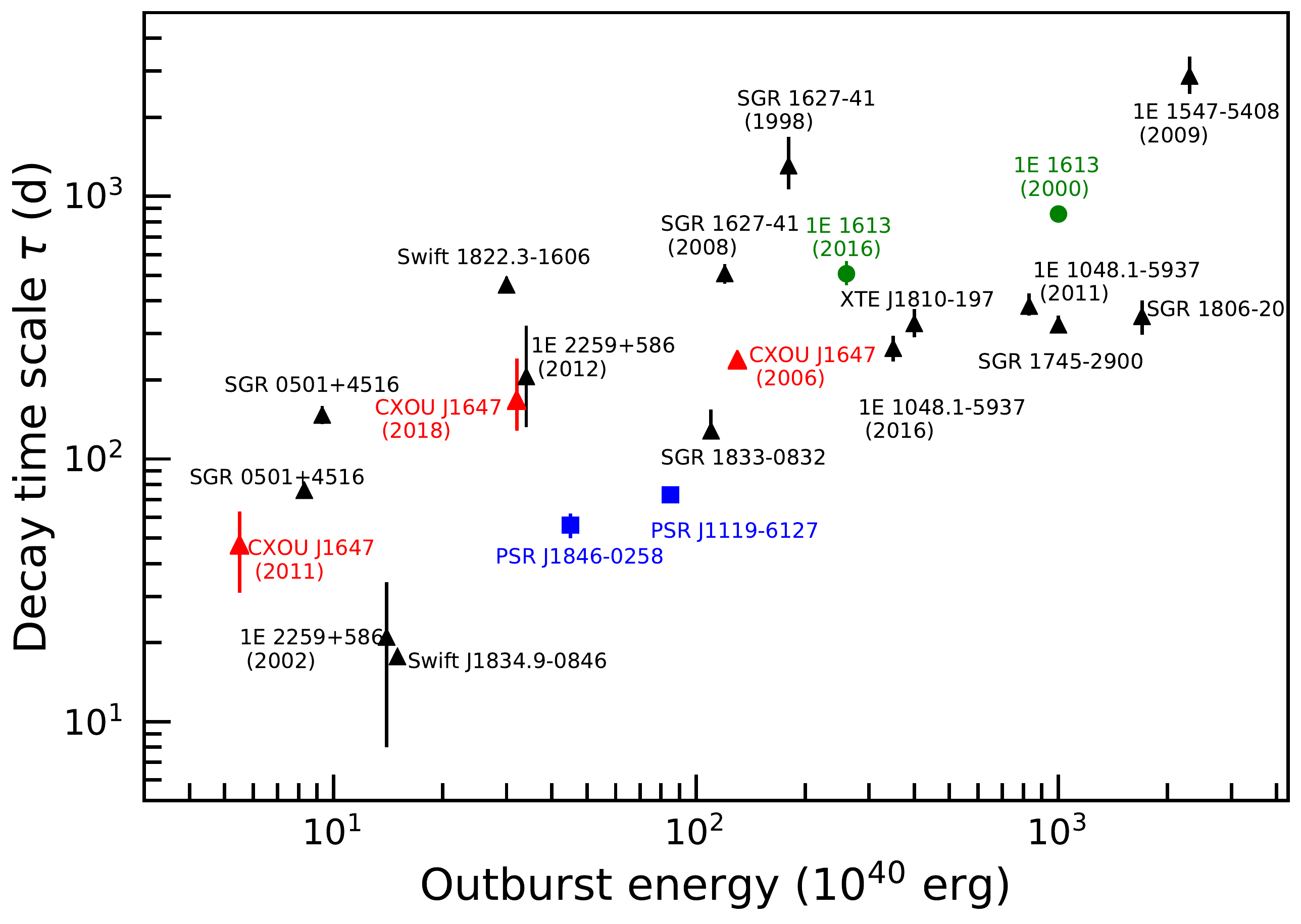}
\end{center}
\caption{Decay time scale (in term of $e$-folding time) as a function of the total energy released for magnetars showing major outbursts. The triangles refer to the ``canonical'' magnetars and in red we highlighted the 2006, 2011 and 2018 outbursts of \srclong. Blue squares indicate the rotation-powered pulsars with high magnetic fields that showed magnetar-like activity. The green circles denote the two outbursts of 1E\,161348--5055, the central source of the supernova remnant RCW\,103. Adapted from \citet{2018MNRAS.474..961C}.}

\label{fig:corr}
\vskip -0.1truecm
\end{figure}

\begin{figure*}
\begin{center}
\includegraphics[scale=0.35]{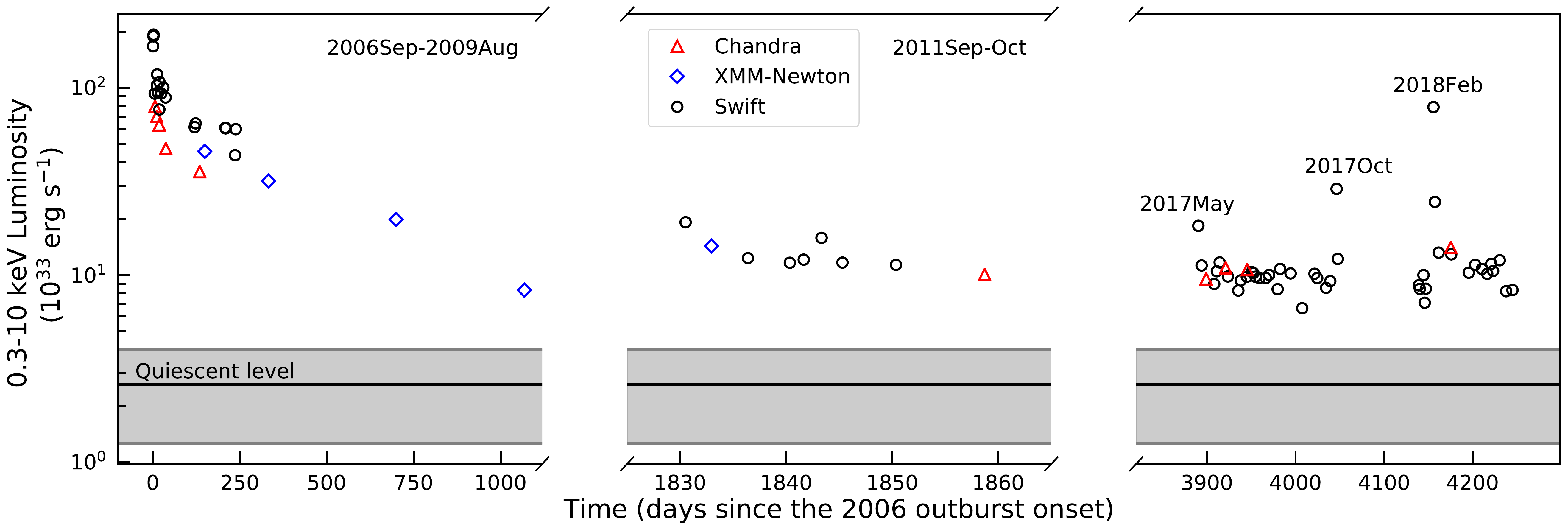}
\end{center}
\caption{
Long-term evolution of the 0.3 -- 10~keV luminosity of \srclong\ from \cxo\ (red triangles), \xmm\ (blue diamonds) and \swift\ (black circles) data acquired between 2006 September and 2018 April. The 2006 outburst onset occurred on 2006 September 21 at 01:34:52 UTC \citep{2006ATel..894....1K}. The solid black line denotes the quiescent level (2.6 $\times$ 10$^{33}$~\lum) derived with a RCS model by  \citet{2013MNRAS.434..123V} from the \xmm\ observation performed on 2006 September 16. The grey area represents the associated uncertainty.}
\label{fig:lumlongterm}
\vskip -0.1truecm
\end{figure*}

\src\ revealed to be a rather prolific magnetar over the past
decade, showing two outbursts in 2006 and 2011; the energy
released in the 2018 February outburst makes this event the second
most powerful recorded so far from this source, with an energy
release about a factor of 3 lower than that in
2006 ($E$ $\sim$ 10$^{42}$~erg), and a factor of $\sim$ 5 larger
than that measured following the 2011 event ($E$ $\sim$ 6 $\times$
10$^{40}$~erg). The 2006 and 2011 outbursts are characterized by a
decay time scale of $\sim$ 240~d and 50~d, respectively; the time
scale of the 2018 event ($\sim$ 170~d) is in between these two
values. This result nicely fits in the correlation between the
total outburst energy and the corresponding decay time scale found
found for magnetars showing major outbursts \citep{2018MNRAS.474..961C},
implying that the longest outbursts are also the most energetic 
ones (see Figure \ref{fig:corr}). Moreover, the
properties of the 2018 February event also follow the
anti-correlation between the quiescent X-ray luminosity and the
outburst luminosity increase, as well as the
correlation between the energy released during the outburst and
the luminosity reached at the outburst onset,for all magnetar 
outbursts by \citet{2018MNRAS.474..961C} (see Figures 3 and 6 of their work).

%

During the entire monitoring campaign, excluding the epochs
close to the peak of the outbursts, \src\ showed a thermal
spectrum well modelled by an absorbed blackbody. The spectra
corresponding to the XRT pointings following the BAT trigger on
2017 October and 2018 February appeared harder, requesting
an additional component such as a power law. The spectral hardening in correspondence
of bursting activity is an ubiquitous property among magnetars
 \citep{2018arXiv180305716E}. As shown in Figure
\ref{fig:bb}, the inferred blackbody temperature attained a rather
high constant value of $\sim$ 0.7~keV over $\sim$ 350~d; 
the corresponding blackbody radius also settled at a constant value of
$\sim$ 0.5~km during the first $\sim$ 160~d.  It then increased in correspondence of the 
bursts, to $\sim$ 1~km and $\sim$ 2~km, and then slowly decaying towards the
pre-outburst value.

It is interesting to compare the present results from
spectral analysis to those relative to previous outburst episodes
from \src. \cite{2010ApJ...722..788A} used a three blackbody
model, comprising an inner hot cap, a surrounding warm ring and
the cooler remaining part of the surface, to reproduce the pulse
profiles of \src\ over a period spanning more than 1000 d,
starting from the first \xmm\ observation after the September 2006
outburst onset. They found that the temperature of the hot cap
decreased with time from $0.7$ keV to $0.45$ keV, when it merged
with the warm region after $\sim 700$ d. The warm region remained
more or less at constant temperature ($\sim 0.45$ keV), with
possibly a slight increase at later times. The cooler blackbody
was fixed at $0.15$ keV. The area of the hot region shrunk as it
cooled, going from an initial $\sim 8\%$ of the entire surface
to zero in $\sim 700$ d, while the area of the warm corona
increased from $\sim 20\%$ to $\sim 30\%$ of the star surface over
the examined time span. The (phase-resolved) spectral analysis by
\cite{2014MNRAS.441.1305R}, relative to the same time span,
provides a similar picture, with a hotter spot which cools and
shrinks in time and a warm region at roughly constant temperature,
although, at variance with the findings of
\cite{2010ApJ...722..788A}, the area of the latter monotonically
decreases in time. Moreover, the two blackbody temperatures
reported by \cite{2014MNRAS.441.1305R} are somewhat higher.

 Regarding the timing properties, the pulse profile shape of \src\
exhibited quite drastic changes during the previous two outbursts, in 2006 and 2011.
From a multi-peaked configuration at the outburst onset, the pulse profile
returned to the quiescent single-peaked structure \citep[see Figure 2 by][]{2014MNRAS.441.1305R}.
In this latest multi-outburst activity period, the pulse profile exhibited two peaks in the \cxo\ observations 
(time resolution $\sim$ 0.44~s), confirming the behaviour registered during the past flaring events.
In our timing analysis we found an estimate for the period and an upper limit for the period derivative, which 
are consistent with the results previously reported in literature \citep{2011ApJ...726...37W, 2013ApJ...763...82A, 2014MNRAS.441.1305R}.\\
 

The mechanism actually responsible for the heating of the star surface
layers in magnetar outbursts is still not well understood. The
onset of an active phase is most likely due to a rearrangement of
the star external magnetic field, due to the transfer of magnetic
helicity from the interior to the magnetosphere, which results in
the twist of a bundle of field lines. Currents flowing along the
twisted field lines hit the star surface and release heat via
Ohmic dissipation. At the same time, the magnetosphere must
untwist to maintain the potential drop necessary to accelerate the
charges. \citet{2009ApJ...703.1044B} discussed the evolution of a twisted
magnetosphere and provided a simple estimate for the luminosity
released by impinging currents

\begin{equation}\label{belolum}
L_{currents}\sim 10^{36}\left(\frac{B}{10^{14}\,
\mathrm{G}}\right) \left(\frac{R}{10^{6}\,
\mathrm{m}}\right) \left(\frac{\cal V}{10^{9}\,
\mathrm{V}}\right) \psi\sin^4\theta_*\ \mathrm{erg/s}\,,
\end{equation}

where $\cal V$ is the potential drop, $\psi$ is the twist angle
($\psi\la\psi_{max}\sim 1\, \mathrm{rad}$) and $\theta_*$ is the
opening angle of the twisted bundle (which is assumed to be
localized around the pole). In the case of \src, taking reference
values in equation \ref{belolum}, $\psi\sim 1\, \mathrm{rad}$ and
$\theta_*\la 0.1\, \mathrm{rad}$ (this follows from the measured
blackbody radius $\sim 0.1$-$1$ km), the luminosity turns out to
be $L_{currents}\la 10^{32}\ \mathrm{erg/s}$. Although non-polar
twists can produce a higher luminosity, the previous value is
about two orders of magnitude below what observed. This implies
that ohmic dissipation of returning currents alone is unlikely to
produce the observed thermal flux in \src. On the other hand, the
predicted evolution timescale of the untwisting magnetosphere,

\begin{equation}\label{belotime}
t_{ev} \sim 15 \left(\frac{B}{10^{14}\,
\mathrm{G}}\right) \left(\frac{R}{10^{6}\,
\mathrm{m}}\right)^2 \left(\frac{\cal V}{10^{9}\,
\mathrm{V}}\right)^{-1} \psi\sin^2\theta_*\ \mathrm{yr}\,,
\end{equation}

turns out to be $\sim 1$ month, quite in agreement with
observations.

Schematically, the global scenario could then be summarized as
follows. Consistently with the expectations of cooling models
\citep{2013MNRAS.434..123V}, most of the star has a relatively low
temperature (0.1 -- 0.2~keV) in its quiescent state. This component
could only be detected in a few cases because of the typically high
absorption. During the evolution, energy and helicity are transferred
from the interior to the magnetosphere until some instability triggers
a global magnetic reconnection. The high temperature (0.7 -- 1.0~keV),
in a very localized component, is likely produced by returning
currents of a bundle hitting on the star surface. The energy released
in the crust is unlikely to cause such a high surface temperature
since the process is not efficient due to neutrino losses and the
spread of the heat wave \citep{2012ApJ...750L...6P}. The origin of the
intermediate component (0.3 -- 0.5~keV), interpreted as a warm ring
around the shrinking central hot spot, is less clear. In most
magnetars, this warm component can survive for a long time (years), in
most cases being even part of the quiescent emission, and being
relatively stable for a decade or more. This does not quite fit in the
purely magnetospheric bundle picture, which should be dissipated
relatively fast (months). Thus, this intermediate component must be
somehow maintained by a continuous energy injection from the
interior. Impulsive energy release in the crust has been
systematically explored in the literature \citep{2014MNRAS.442.3484K,
  2018arXiv180706855C} and may be part of the solution, although it
also has some problems. In particular, multi-D models predict the
widening of the warm spot, which is not usually observed. A new
interesting idea has recently been proposed by
\citet{2018arXiv180709021A} who analyzed the coupled evolution of the
interior of the star and of a force-free magnetosphere (see also
\citet{2017MNRAS.472.3914A}). They have estimated the effect that the
currents going through the envelope would have on the surface
temperature and found that the last $\approx 1$ meter below the
surface can be kept at a high temperature in the quasi-stationary
regime. Basically, they found that, to close the global current
circuit maintaining the twisted magnetosphere on long timescales,
currents must go through the low density region between the crust and
the exterior, where the electrical resistivity is highest. Releasing
energy by Ohmic dissipation in a thin layer of a few meters is very
efficient, and the small volume implied requires much less energy to
raise the surface temperature to observed values than releasing energy
deep in the star crust.\\


We will continue monitoring the \src\ with \swift\ XRT to follow
it while recovering its quiescent phase or possibly stabilizing to a new quiescent state, unveil any
significant spectral and/or timing evolution, and refine the
outburst energetics and decay time scale.

%

\section*{Acknowledgements}
The results reported in this paper are based on observations obtained with \cxo, \nus, \swift\ and \INT. The \nus\ mission is a project led by the Californian Institute of Technology, managed by the the Jet Propulsion Laboratory and funded by NASA. \swift\ is a NASA mission with participation of the Italian Space Agency and the UK Space Agency. We made use of the software provided by the {\em Chandra X-ray Center} (CXC) in the application package {\sc CIAO}. \INT\ IBIS/ISGRI has been realized and maintained in flight by CEA-Saclay/Irfu with the support of CNES.  
AB, NR and PE are supported by an NWO Vidi Grant (PI: Rea). NR is also supported by grants AYA2015-71042-P and SGR 2014-1073.
PE acknowledges funding in the framework of the project ``Understanding the x-ray variabLe and Transient Sky'' (ULTraS), ASI-INAF contract N. 2017-14-H.0. JAP acknowledges support by the Spanish MINECO/FEDER grant AYA2015-66899-C2-2-P, and the grant of Generalitat Valenciana PROMETEOII-2014-069. FCZ is supported by grants AYA2015-71042-P and SGR 2014-1073. DG acknowledges the financial support of the UnivEarthS Labex program at Sorbonne Paris Cite\'e (ANR-10-LABX- 0023 and ANR-11-IDEX-0005-02). We thank the referee for his comments and the COST Action PHAROS (CA16214) for partial support.

\bibliographystyle{mnras}
\bibliography{reference}

\end{document}